\documentclass[natbib]{svjour3}             
\smartqed  
\usepackage{graphicx}
\newcommand{\be}{\begin{eqnarray}}
\newcommand{\ee}{\end{eqnarray}}
\newcommand{\lsim}{\;\raise0.3ex\hbox{$<$\kern-0.75em\raise-1.1ex\hbox{$\sim$}}\;}
\newcommand{\gsim}{\;\raise0.3ex\hbox{$>$\kern-0.75em\raise-1.1ex\hbox{$\sim$}}\;}

\newcommand{\alf}{Alfv\'en}
\newcommand{\growthrate}{\gamma_\mathrm{CR}}
\newcommand{\mm}{$\mu$m}
\newcommand{\ecss}{erg cm$^{-2}$ s$^{-1}$ sr$^{-1}$ polariz$^{-1}$ }
\newcommand{\dd}{\phantom{.5}}
\def\lsim{\;\raise0.3ex\hbox{$<$\kern-0.75em\raise-1.1ex\hbox{$\sim$}}\;}
\def\gsim{\;\raise0.3ex\hbox{$>$\kern-0.75em\raise-1.1ex\hbox{$\sim$}}\;}

\def\cmc{\rm ~cm^{-3}}

\def\kms{\rm ~km~s^{-1}}

\def \kms {\rm ~km~s^{-1}}

\def\alf{Alfv\'en~}
\def\mm{$\mu$m}
\def\ecss{erg cm$^{-2}$ s$^{-1}$ sr$^{-1}$ polariz$^{-1}$ }

\newcommand{\heatpar}{\alpha_H}
\newcommand{\Beff}{B_\mathrm{eff}}
\newcommand{\rgone}{r_\mathrm{g1}}
\usepackage{color}
\usepackage{ulem}

\def\dd{\phantom{.5}}

\begin{document}

\title{Microphysics of partly ionized flows with collisionless shocks
}
\title{Collisionless shocks in partly ionized plasma with cosmic rays:
microphysics of non-thermal components}


\titlerunning{Shocks in partly ionized plasma with cosmic rays}        

\author{A.M. Bykov \and  M.A. Malkov \and \\ J.C. Raymond \and A.M. Krassilchtchikov  \and
            A.E. Vladimirov}

\authorrunning{A.M.~Bykov, M.A.~Malkov, J.C.~Raymond et al.} 

\institute{A.M.~Bykov \at A.F.~Ioffe Institute for Physics and
Technology, 194021, St.Petersburg, Russia \\
\email{byk@astro.ioffe.ru} \and
M.A.~Malkov \at University of California, San Diego, La Jolla,
California 92093, USA \\ \email{mmalkov@ucsd.edu} \and J.C.~Raymond \at
Center for Astrophysics, 60 Garden St, Cambridge, MA 02138, USA \\
\email{jraymond@cfa.harvard.edu} \and A.M.~Krassilchtchikov \at
A.F.~Ioffe Institute for Physics and Technology, 194021,
St.Petersburg, Russia \\ \email{kra@astro.ioffe.ru} \and
A.E.~Vladimirov \at Stanford University, Stanford, CA 94305, USA
\email{avladim@stanford.edu} }


\date{Received: date / Accepted: date}

\maketitle
\begin{abstract}
In this review we discuss some observational aspects and theoretical
models of astrophysical collisionless shocks in partly ionized
plasma with the presence of non-thermal components. A specific
feature of fast strong collisionless shocks is their ability to
accelerate energetic particles that can modify the shock upstream
flow and form the shock precursors. We discuss the effects of
energetic particle acceleration and associated magnetic field
amplification and decay in the extended shock precursors on the line
and continuum multi-wavelength emission spectra of the shocks. Both
Balmer-type and radiative astrophysical shocks are discussed in
connection to supernova remnants interacting with partially neutral
clouds. Quantitative models described in the review predict a number
of observable line-like emission features that can be used to reveal
the physical state of the matter in the shock precursors and the
character of nonthermal processes in the shocks. Implications of
recent progress of gamma-ray observations of supernova remnants in
molecular clouds are highlighted. 
\end{abstract}


\section{Introduction}
\label{intro}

Plasma flows with collisionless shocks are found in a number of
energetic space objects, such as supernova remnants (SNRs)
interacting with atomic or molecular clouds
\citep{McKee_Hollenbach_80, Draine_McKee_93,heng10}, Herbig-Haro
objects \citep{HRH_87, Giannini_ea_08, Tesileanu_ea_09}, winds from
protostars and young stellar objects \citep{Massi_ea_08}, recurrent
novae \citep{Evans_ea_07, Bode_ea_07, Tatischeff_Hernanz_07}, and
also in accretion flows in the vicinities of galactic nuclei
\citep{Farage_ea_10}.

Already in the early studies of \citet{Raymond_79} and \citet{Shull_McKee_79}
the ionization state of the pre-shock flow was recognized as an important
 feedback parameter that influences the dynamics of the flow, and hence,
the spectrum of continuum and line emission produced in the post-shock zone.
While \citet{Raymond_79} took the pre-shock ionization state as a free parameter,
\citet{Shull_McKee_79} included the ionizing flux from the post-shock to
obtain a pre-shock ionization state self-consistently with the temperature and
ionization profiles in the post-shock.  \citet{DS_96} furthered the calculation
of the precursor photoionization, and presented emission line
spectra assuming an equilibrium with the ionizing flux from the shock.

Recent observations of forward shocks in galactic SNRs indicate a
substantial role of nonthermal components in the energy budgets of
post-shock flows \citep[e.g.,][]{helderea12}. While the fraction of
the energy dissipated in the shock itself may be smaller for slower
radiative shocks,  the enhanced radiative cooling in their
post-shocks leads to stronger compression so that cosmic ray and
magnetic pressure can dominate over thermal pressure in the zones
where most of the observed optical and IR emission is produced.
There is also a growing consensus on that in both parallel and
perpendicular shocks the diffusing cosmic rays may generate unstable
plasma currents and turbulence in the pre-shock zone, thus allowing
for efficient particle acceleration, shock modification, and
substantial enhancement of the total compression ratio
\citep{blandfordeichler,je91,MDru01,bell04,ber12}.
\citet{Raymond_ea_88} analyzed a shock in the Cygnus Loop SNR and
found that the nonthermal pressure exceeds the thermal pressure by
an order of magnitude in the zones where the [S II] lines are
formed, though they were unable to distinguish between magnetic and
cosmic ray contributions. Neutral particles could affect the
processes of particle acceleration \citep[e.g.,][]{Draine_McKee_93,
ddk96,MDS05,blasiea12,Inoue12,ohira12, helderea12,morlinoea12},
magnetic field amplification and plasma heating in the upstream
region \citep[e.g.,][]{bt05,revilleea07}.

In this paper we review physical mechanisms governing evolution of
supersonic and superalfvenic  flows with collisionless shocks in
partially ionized plasmas with non-thermal components.
Section~\ref{sec: photo} describes the photoionization precursors
that help determine the neutral fraction of the pre-shock gas. In
Section~\ref{sec:interaction} H$\alpha$ signatures of interaction of
neutral particles with fast shocks are depicted. In
Section~\ref{sec:tp} we discuss specific plasma heating processes
due to dissipation of CR-driven magnetic fluctuations in the shock
precursor. A consistent pre-shock ionization structure is given in
Section~\ref{sec:1} with an aim to emphasize processes that
critically influence optical and IR spectra emitted from the flows.
The escape of accelerated energetic particles may affect the gas
compression in the shock upstream and plasma temperature in the
downstream resulting in formation of radiative shocks of velocities
that may well exceed that of the adiabatic shocks. We discuss the
line spectra of radiative shocks modified by CR acceleration effect
in Section~\ref{sec:1}. Specific features of CR acceleration in
supernova remnants interacting with molecular clouds are discussed
in Section~\ref{sec:breaks} and their gamma-ray emission spectra are
outlined in Section~\ref{sect:gammaSNR}. A brief summary is given in
Section~\ref{sect:summ}.

\section{Photoionization Precursors}
\label{sec: photo}
The neutral fraction in the pre-shock gas often depends on
photoionization by radiation from the shock itself.  There are
two regimes,  non-radiative shocks are those in which gas is
heated at the shock but does not have time to radiatively cool,
while radiative shocks radiatively cool from the post-shock
temperature, efficiently converting their thermal energy into
radiation. Naturally, the latter produce more ionizing photons
if the post-shock temperature exceeds about 150,000 K.  The
scale of the photoionization precursor is given by the photoabsorption
cross section of hydrogen and the neutral density in the pre-shock
gas, so it is typically $10^{18}$ or $10^{19}$ cm.  In supernova
remnants, shocks faster than 300 km/s are typically nonradiative,
while slower shocks are generally radiative.

Non-radiative shocks produce some ionizing photons because some of
the atoms passing through the shock are excited before they are
ionized in the hot post-shock plasma.  In particular, at high
temperatures each He atom produces a few photons in the He I 58.4 nm
and He II 30.4 nm lines.  Since He makes up about 10\% of the gas,
the pre-shock region will be by 30\% to 40\% ionized by these
photons. EUV emission from the hot post-shock gas can further ionize
the plasma. The photons are relatively energetic, so they heat the
gas to around 16,000 K. Emission from such precursors outside
Tycho's SNR and N132D were reported by \cite{ghavamian00} and
\cite{morse_96}.

As mentioned above, and as described more fully in
Section~\ref{sec:1}, radiative shocks faster than about 100 km/s
produce large fluxes of ionizing photons, and those above about 120
km/s produce enough photons to fully ionize H in the pre-shock gas.
However, it should be kept in mind that the equilibrium between
pre-shock ionization state and ionizing flux from the shock, as
assumed in \cite{Shull_McKee_79, DS_96}, does not always hold.  In
particular, an SNR shock can evolve more rapidly than the
recombination time in the pre-shock gas, or the transverse scale of
the shock may be smaller than the upstream photon mean free path, as
it is in a Herbig-Haro object \citep{Raymond_88a}.

\section{Interaction of Neutrals with Fast Shocks: H$\alpha$ Signatures}
\label{sec:interaction}

\begin{figure}
\hspace*{0.4in}
  \includegraphics[width=0.8\textwidth]{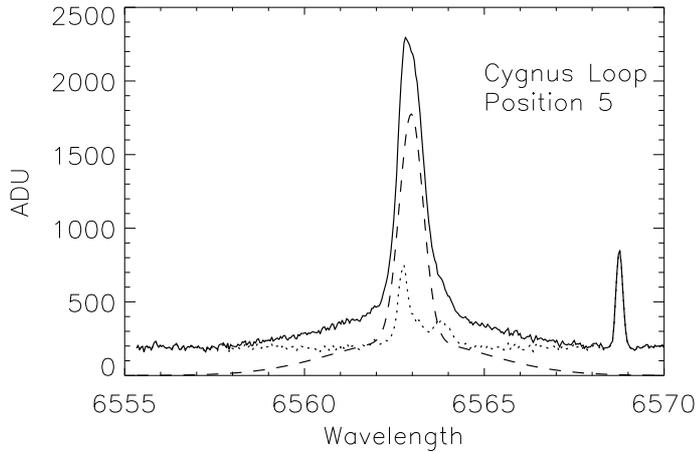}
\caption{H$\alpha$ profile of a 350 km/s shock at the outer edge of
the Cygnus Loop, obtained with the HECTOCHELLE spectrograph on the
MMT.  The narow component is 30 km/s wide (FWHM), and the broad
component is 225 km/s wide.  Full line is the observation, dashed
line is the H$\alpha$ from the filament, and dotted line is the sky
background. The feature near 6569\AA\/ is an airglow line.}
\label{cygloop}       
\end{figure}

Neutral H atoms are not affected by the electromagnetic fields or
plasma turbulence in a collisionless shock.  They pass freely
through the shock transition, whose thickness is given by the ion
inertial length or the proton gyroradius, which are typically $\sim~10^8$~cm
in the interstellar medium. However, they become ionized in
a hot downstream plasma but, because of the low ISM densities,
they travel $\sim 10^{14}$-$10^{15}$ cm downstream before it
happens.  This is smaller than electron-proton thermal equilibration scales,
and therefore, the emission from neutrals may contain information about
the post-shock conditions of electrons and ions before relaxation.

The hydrogen atoms are ionized either by collisions with protons and
electrons or by a charge exchange with protons.  The latter process
produces a population of neutrals with a velocity distribution
similar to that of the post-shock protons and, when they are
excited, they produce broad Lyman and Balmer emission lines whose
velocity widths are comparable to the shock speed \citep{cr78,
raymond91, heng10, france11}. Neutrals that are excited before the charge
transfer produce a narrow component whose width is given by the
pre-shock temperature (see Figure~\ref{cygloop}).  The
broad-to-narrow intensity ratio indicates the electron temperature,
while the broad component width directly measures the proton
temperature.  Thus one can determine the electron-ion temperature
ratio at the shock (Ghavamian and Schwarz this volume).
Consequently, the H$\alpha$ profile provides both the post-shock
temperature and $T_e/T_p$ ratio.
These quantities can be used  to determine the shock speed
\citep{smith94, hester, Ghavamian_ea_01, vanAdelsberg_ea_08}. Under
the assumption that only a small fraction of the shock energy goes
into cosmic rays, the shock speed can be determined from the
post-shock proton and electron temperature using momentum and energy
conservation laws. However, if cosmic ray acceleration is efficient,
then $T_e$ and $T_p$ combined with a model of efficient cosmic ray
acceleration can yield a shock velocity \citep{Helder_ea_09} though
there is some ambiguity due to a contribution to the narrow
component from a shock precursor (see
Section~\ref{sect_precursors}).

The Balmer line profiles can also indicate more exotic
plasma processes.  First, if the post-shock proton velocity
distribution is not Maxwellian, it will leave an imprint on the
broad component line profile.  Second, the strength and width of the
narrow component can indicate heating and excitation in a shock
precursor.  We discuss signatures of  these departures from
a simple shock picture.

\subsection{Non-Maxwellian Velocity Distributions}

If cosmic rays are accelerated in SNR shocks, the particle velocity
distribution is manifestly non-Maxwellian, taking the form of a
Maxwellian core with a power law tail.  Such distributions are
conveniently parameterized with the $\kappa$ function
\citep{pierrard_lazar10}.  The $\kappa$
distribution approaches a Maxwellian as $\kappa$ approaches
infinity, while the power law tail becomes harder and includes a
larger fraction of the particles as $\kappa$ approaches its lower
limit of 1.5.

A different sort of departure from a Maxwellian can arise because of
the neutral atoms themselves.  When a neutral becomes ionized, it is
moving at 3/4 the shock speed with respect to the downstream plasma
and magnetic field (assuming a shock compression factor of 4).  It
therefore behaves like a pickup ion in the solar wind
\citep{gloecklergeiss}. Initially, all these newly-formed protons
have the same velocity with respect to plasma in which they are
immersed.  This velocity separates into a gyration speed around the
magnetic field plus a component along the field, and since all the
particles  have the same initial velocity (assuming that the
shock speed is large compared to the pre-shock thermal speed), they
form a ring beam in velocity space.  That distribution is unstable,
and it rapidly evolves into a bispherical shell in velocity space, a
lens-like shape that depends on the ratio of the Alfv\'{e}n speed to
the initial particle speed \citep{williamszank}.  The protons in the
bispherical shell can then experience charge transfer to produce
observable neutrals. The process of ionization, interaction with the
magnetic field, and subsequent re-neutralization is the same
sequence of events that produce the ``IBEX ribbon" seen in H atoms
from beyond the heliopause \citep{mccomas09}.

The H$\alpha$ profiles produced by the pickup ion process were
computed by \cite{raymond08}. The neutral fraction in the pre-shock
gas must be substantial in order to produce the Balmer line
filaments seen at the edges of some supernova remnants and
pulsar-wind nebulae, and therefore the neutrals can have a significant
effect on the shock structure.  \cite{Heng_ea_07} computed the gradual
transition over the charge transfer length scale, but found little
effect on the overall dynamics.  \cite{ohira09} examined the
streaming instabilities created by the relative motion of the
post-shock plasma and the protons created downstream (this effect is stronger in
parallel shocks) and predicted substantial amplification
of the magnetic field. \cite{raymond08} calculated the wave energy
produced as the ring beam relaxes to a bispherical distribution, and
showed that it could significantly affect the electron temperature
if the waves couple efficiently to electrons.  \cite{ohira10}
considered the modification of the shock structure, in particular
weakening of the subshock, and the effects on the cosmic ray
spectrum.

The observational evidence for any departure of the proton
distribution from a Maxwellian is still ambiguous. For nearly all
the observed H$\alpha$ broad components Gaussian fits are adequate,
but in general the lines are so faint that
even the determination of the line width is subject to large statistical uncertainty.
A very deep exposure of a bright knot in Tycho's SNR yielded
the only profile that shows a clear non-Gaussian broad component,
but there are several possible interpretations \citep{raymond10}.
The departure could be attributed to a power law tail, however,
the slope of this tail is very hard, which indicates extremely
efficient particle acceleration.  The profile could also be matched
by the combination of a pickup ion component with an ordinary
Maxwellian component, though this requires a fairly high pre-shock
neutral fraction.  A third possibility is that some of the neutrals in the  shock precursor
aquire a kinetic temperature about 1/2 of the
post-shock temperature.  Clearly, this implies a precursor to be both
hot and thick.  Finally, one could produce the observed profile by
adding contributions from separate shocks along the line of sight. However, this possibility
would require quite a large velocity difference, and
therefore a significant density contrast between the two regions.  At higher
shock speeds, the velocity dependence of the charge transfer cross
section can distort the profile \citep{heng07a, Heng_ea_07}, but that is
unlikely to account for the Tycho observations.

\subsection{Shock Precursors}
\label{sect_precursors}

The precursor of a collisionless shock wave is a region upstream of the shock
transition in which the plasma conditions (velocity, density, temperature,
magnetic fields, ionization state) are affected by photons or superthermal
particles streaming ahead of the shock front.

Shock wave precursors can be produced by ionizing photons emerging from
the downstream region or by broad component neutrals that leak back through the shock.
Additionally, in shocks that produce accelerated super-thermal charged particles,
precursors can be produced as a consequence of this acceleration process.
We first consider the
observational evidence for precursors, then the physics of the three
mechanisms and the implications for system parameters.

The first evidence for precursors to SNR shocks came from the
observation that the widths of the H$\alpha$ narrow components in
several shocks were unexpectedly large.  \cite{smith94} measured
narrow component widths of 25-58 km/s in four LMC Balmer-dominated
remnants, and \cite{hester} found a 28-35 km/s width in a Cygnus
Loop shock.  These line widths correspond to temperatures of about
(2--7)$\times$ 10$^4$ K, and in static equilibrium, hydrogen is
fully ionized at those temperatures.  Thus \cite{smith94} and
\cite{hester} concluded that the gas must be heated in a narrow
region ahead of the shock. The region must be thick enough so that
charge transfer can heat the neutrals, but thin enough that the
neutrals do not become ionized.  Subsequent observations have
revealed line widths of 44 km/s  in Tycho's SNR \citep{ghavamian00}
and 30-42 km/s in RCW 86 and Kepler's SNR, with only SN~1006 showing
a narrow component width compatible with a temperature of 10,000 K
and a significant neutral fraction \citep{sollerman}.
\cite{nikolic_ea_13} have presented Integral Field Unit spectra of a
section of the H$\alpha$ filament in SN~1006, and they have found
that a precursor makes a substantial contribution to the narrow
component of H$\alpha$.

The precursor thicknesses must be about 1$^\prime$$^\prime$ in order
for the neutrals to be heated, and they have been spatially resolved
in the Cygnus Loop \citep{fesenitoh, hester} and in Tycho's SNR
\citep{lee07, lee10}.

\subsection{Physical Interpretation of Shock Precursors}
 Three ideas have been put forward to explain observed shock
precursors, a photoionization precursor, a cosmic ray precursor, or
a precursor created by fast neutrals leaking from the post-shock
region out ahead of the shock.

The photoionization precursor heats the electrons, and it has a very
large length scale.  In non-radiative shocks, each He atom passing
through the shock produces about 2 He II $\lambda$304 photons and 1
to 2 He I $\lambda$584 photons. These photons can ionize about 30\%
of the pre-shock hydrogen, and because the photons are relatively
energetic they deposit considerable heat in the electrons. Such a
precursor has been reported ahead of Tycho's supernova remnant
\citep{ghavamian00}. However, because they heat the electrons and
extend over large length scales, photoionization precursors do not
explain the observations discussed in Section~\ref{sect_precursors}.

A  precursor is an integral part of diffusive shock acceleration
models, in which an MHD turbulence scatters energetic particles
moving away from the shock back towards it \citep{blandfordeichler}.
Its thickness is given by $\kappa_{CR} / V_s$, where $\kappa_{CR}$
is the cosmic ray diffusion coefficient. While $\kappa_{CR}$ is of
order $10^{28}$ cm$^2$/s in the ISM, it must be somewhere closer to
the Bohm limit near the shock in order for particles to reach high
energies, perhaps $10^{24}$ cm$^2$/s. That implies a precursor
thickness of order $10^{16}$ cm, or about 1$^\prime$$^\prime$ for
nearby SNRs. The heating in such a precursor can occur by
dissipation of the turbulence itself. It will be especially strong
if neutrals are present to provide ion-neutral wave damping, which
can limit the wave intensity and therefore the highest particle
energies reached \citep{Draine_McKee_93, ddk96}. The heating will
also be strongest in shocks that accelerate cosmic rays efficiently.
If the precursor heating is strong, it can reduce the Mach number of
the subshock, changing the cosmic ray spectrum and pressure
\citep[see, e.g.,][]{vbe08,Wagner_ea_09, vink10}.

Several theoretical models have considered the effects of
neutral-ion collisions on the dynamics and temperature of the
precursor. \cite{boularescox} computed the heating and ionization in
cosmic ray precursors for the modest shock speeds in the Cygnus
Loop. \cite{ohira10} considered neutrals interacting in the
precursor, treating them as pickup ions. The pickup ions can
strongly heat the plasma and affect the jump conditions at the
subshock, and because of their high velocities they are
preferentially injected into the diffusive acceleration process.
\cite{raymondea11} computed the temperature and density structures
of precursors and the resulting H$\alpha$ emission.  If cosmic ray
acceleration is efficient, the H$\alpha$ profiles are strongly
modified, both in broadening the narrow component and decreasing the
broad-to-narrow intensity ratio. The H$\alpha$ profiles reported so
far are consistent with moderate ratios of cosmic ray pressure to
gas pressure, but not with ratios above about 0.4.  Most recently,
\citet{morlino_ea_13} have included a parameterized form of wave
damping along with wave growth due to the cosmic ray pressure
gradient for a more self-consistent treatment of the cosmic ray
spectrum in the presence of neutrals and the resulting H$\alpha$
profiles. They also find that an intermediate velocity width
component can arise, similar to the complex profiles predicted in
\cite{raymondea11}.

The other suggested mechanism for producing a precursor is leakage
of some of the broad component neutrals out through the shock front.
These neutrals have the thermal speeds of post-shock protons and a
bulk speed of $V_S$/4 away from the shock.  A significant fraction
of them, of order 10\%, can overtake the shock, and of course they
pass through \citep{smith94, hester}.  They carry a substantial
amount of energy, but it is unclear how much of that energy heats
the precursor plasma. \cite{limraga} found that they became ionized
to form a beam of fast protons, but that the protons are swept back
through the shock without transferring much energy to the bulk
plasma. \cite{ohira12} constructed four-fluid models and showed that
the leaking particles decelerate the upstream flow and affect the
subshock compression ratio. \cite{blasiea12} computed the energy and
momentum exchange between neutrals and ions on both sides of the
subshock, using a procedure like that of \cite{Heng_ea_07} to
compute the neutral velocity distribution and the flux of neutrals
from the post-shock region to the precursor.  Then they used the
resulting precursor structure to compute spectral slopes of
accelerated particles. The authors accounted  for neutrals
ionization by  protons only, though photo ionization processes can
affect the ionization states for some shock parameter space.
\cite{morlinoea12} used the kinetic simulations of \cite{blasiea12}
to calculate the H$\alpha$ emission produced in the precursor. Its
intensity depends on the poorly constrained efficiency of
electron-ion equilibration, since the heat is deposited in the ions,
but the electrons excite the line. They find that the profiles can
deviate strongly from Gaussian and that in some circumstances an
intermediate width component can arise. They suggest that this could
explain profiles observed in Tycho's SNR. \citet{morlino_ea_13} have
extended this work by including the cosmic rays along with the
neutrals that escape upstream through the shock.

\section{CR precursor heating and the post-shock temperature}
\label{sec:tp}
\begin{figure}
\hspace*{0.4in}
  \includegraphics[width=0.8\textwidth]{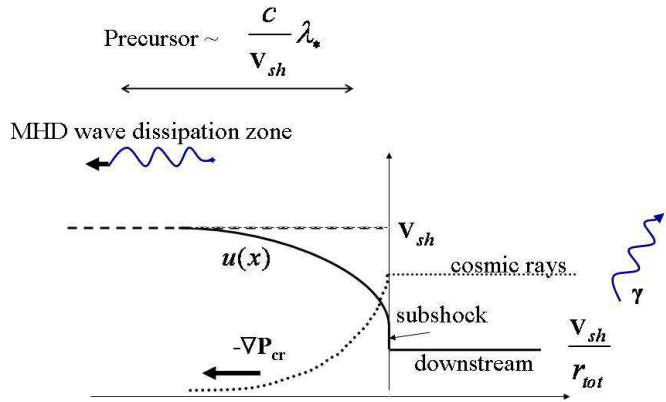}
  \caption{A schematic view of a CR-modified shock
propating to the right.}
\label{sketch}       
\end{figure}

Models of collisionless shocks with large sonic and Alfv\'{e}nic Mach numbers
(${\cal M}_{\rm s}>> 1$ and Alfv\'{e}nic ${\cal M}_{\rm a}>> 1$) show that,
through the first-order Fermi acceleration mechanism,
a small minority of
particles could gain a disproportionate share of the energy and
populate the high energy tail of particle distribution. The energetic
particles can penetrate far into the shock upstream gas, to create
an extended shock precursor illustrated in Figure~\ref{sketch}. The
cold gas in the shock precursor is decelerated and pre-heated by
fluctuating magnetic field dissipation on a scale that is about
$c/v_{\rm sh}$ times larger than a mean free path of an energetic
particle $\lambda_{\ast}$. Shocks in collisionless supersonic flows
produce a complex multi-scale structure of the relaxation region
with an extended precursor and sub-shock of a modest sonic Mach
number ${\cal M}_{\rm sub} \sim$ 3.

This section discusses theoretical models of strong collisionless shocks
and their implications for the observations of shock precursors discussed
in Section~\ref{sect_precursors}.

\subsection{The structure of a non-linear shock precursor
in the presence of self-generated MHD turbulence}

Consider a strong, plane-parallel collisionless shock in a plasma of
finite $\beta = {\cal M}^2_{\rm a}/{\cal M}^2_{\rm s}$ ($\beta$ is
the ratio of plasma pressure to magnetic pressure). The distribution
function of nonthermal particles and the bulk flow profile in the
shock upstream region are sensitive to both the total upstream
compression ratio $r_{\rm tot}$ and the subshock Mach number ${\cal
M}_{\rm sub}$. Direct numerical simulations of the CR-modified shock
by particle-in-cell technique are nonfeasible by now because of the
wide dynamical range of the simulation that requires extreme
computing resources. Nevertheless, an approximate iterative approach
(e.g., within the Monte Carlo model discussed in \citet{vbe08} or
semi-analytical kinetic models developed by \citet{malkov97,ab06})
can be used to derive the steady-state distribution function
consistent with the shock compression. These approximate models
assume some diffusion model and parameterize the microphysical
processes of magnetic field amplification and plasma heating.  The
exact calculation of the CR escape flux $Q_{\rm esc}$ that
determines the total upstream compression ratio $r_{\rm tot}$ can be
performed only in fully nonlinear self-consistent simulations. The
Monte Carlo model of \citet{vbe08} describes the escape of particles
from the shock with an assumed free escape boundary far upstream of
the shock. The distance to the free escape boundary is a free
parameter of the simulation that controls the maximum energy of
accelerated particles and the escaping CR flux (see \cite{ebj96}
for more details of this method).


\begin{figure}
\hspace*{0.4in}
  \includegraphics[width=0.8\textwidth]{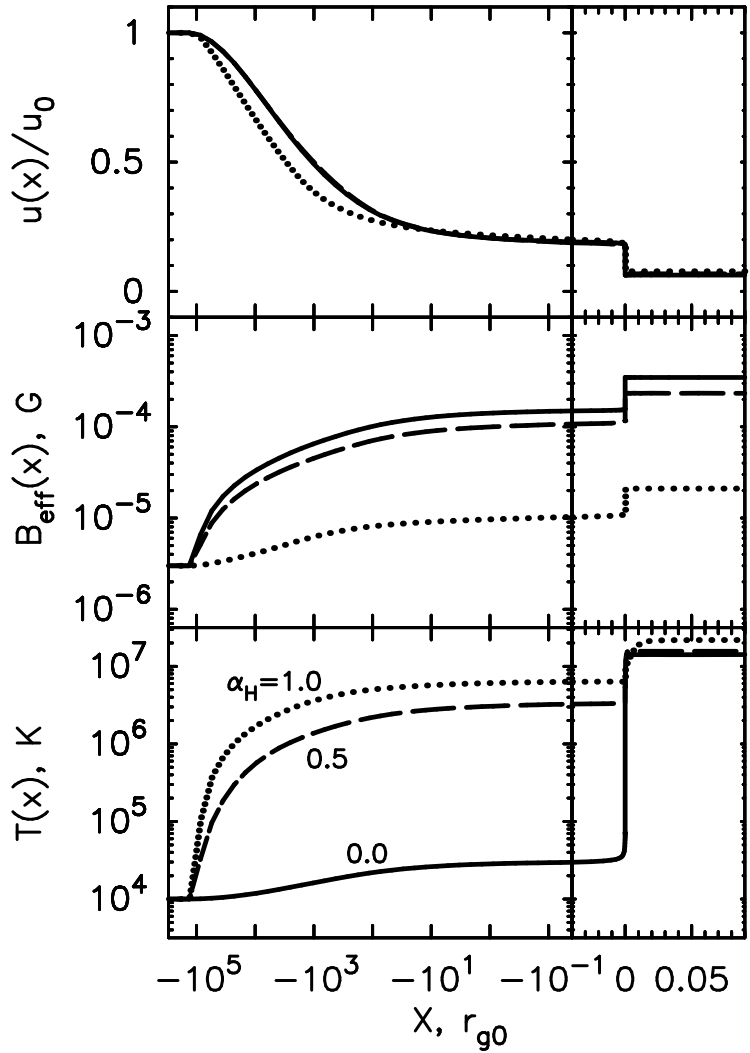}
  \caption{The profile of a CR-modified shock simulated with
Monte-Carlo nonlinear DSA model by \citet{vbe08}. The solid, dashed
and dotted lines correspond, respectively, to a fraction, $\heatpar
= 0$, $0.5$ and $1.0$, of the magnetic energy that goes into heating of
the thermal gas. The plotted quantities are the bulk flow speed
$u(x)$, the effective amplified magnetic field $\Beff(x)$ and the
thermal gas temperature $T(x)$. The shock is located at $x=0$.}
\label{vbe}       
\end{figure}

The role of accelerated particles in the shock precursor depends on the
poorly known mechanisms with which energetic particles can transfer energy to the thermal pool.
However, in order to numerically estimate the impact of such a process,
\citet{vbe08} parameterize the rate of plasma heating upstream.
In order to do that, they assume that CR particles produce, through the resonant CR streaming instability,
strong fluctuations of magnetic fields upstream of the shock.
Another assumption in the model is that these fluctuations are immediately dissipated into the heat of the thermal gas,
and that the rate of turbulence dissipation is proportional to the rate of turbulence generation.
The dimensionless parameter $\alpha_H$ in the model is the ratio of dissipation to generation rate.
Figure~\ref{vbe} adopted from \cite{vbe08} shows the self-consistently calculated flow speed,
effective magnetic field and the temperature
in the precursor of a strong shock with turbulence dissipation for several values of $\alpha_H$.
This work demonstrated that even a small rate of turbulence dissipation can significantly increase
the gas temperature in the precursor, and thus increase the rate of injection of thermal particles.
At the same time, the spectrum of high energy particles is only modestly affected by dissipation in these simulations.

In a follow-up work of \citet{vbe09} the effect of the microphysics of the fluctuating magnetic
fields in the shock precursor was modeled in a more physically realistic manner.
That work used the balance equation of the turbulence spectral energy density
\begin{equation}
\label{eq_turb_evol} \frac{\partial W}{\partial t} + \mathbf{u}\,
\nabla_\mathbf{r} W +  \nabla_\mathbf{k} \mathbf{\Pi} =
S(x,\mathbf{k}) - L_{\rm turb}.
\end{equation}
 Here $L_{\rm turb}(x,\mathbf{k})$ represents the dissipation of turbulence,
the turbulence energy injection rate is
 $S(x,\mathbf{k}) = \gamma (x,\mathbf{k}) W(x,\mathbf{k},t)$,
and $\gamma (x,\mathbf{k})$ is the rate of wave energy amplification
by the CR-driven instabilities \citep[see for a
discussion][]{ber12,schureea12}. The fastest short-wavelength
CR-driven instability studied by \citet{bell04} has the following growth rate
for the wavevector along the mean magnetic field:
\begin{equation}
\label{bell_increment} \growthrate =  2 v_A
k_{\parallel}\,\sqrt{\displaystyle\frac{k_c}{k_{\parallel}}-1},
\quad
  \mathrm{for} \quad 1/\rgone < k < k_c, \label{eq:turbcascade}
\end{equation}
where $v_A(x)=B_0/\sqrt{4 \pi \rho(x)}$ is the \alf\ speed, $c$ is
the speed of light, $B_0$ is the far upstream magnetic field
directed towards the shock normal, $\rho(x)$ is the thermal plasma
mass density, $\rgone(x)$ is the gyroradius of the least energetic
current generating CR, the critical wavenumber $k_c(x) = 4\pi
j_d(x)/(c B_0)$, and the local diffusive electric current of CRs
responsible for the instability, $j_d(x)$, is determined via a
Monte Carlo simulation.  The turbulence energy density flux
$\mathbf{\Pi}(\mathbf{r},\mathbf{k},t)$  describes anisotropic
cascading, i.e., the transfer of turbulence energy from long to
short wavelengths (Kolmogorov type cascade) as well as the inverse
cascade \citep[see, e.g.,][]{mya75,zm90,verma04,zhou10}.

Modeling the balance, spectral distribution, and dissipation
of turbulence energy with equation (\ref{eq_turb_evol})
allows us to construct more physically realistic models of shock precursors
and account for the Balmer line observations discussed in Section~\ref{sec:interaction}.
This equation must be supplemented by theoretical models
of the growth, dissipation and spectral flux of MHD turbulence produced by CRs.
Sections~\ref{sec:neutr} and \ref{sec:heating} discuss some of the ongoing work in this area.

\subsection{MHD turbulence damping in partially ionized plasmas}
\label{sec:neutr} Ion-neutral collisions may dominate the frictional
damping of strong magnetic oscillations in a cold photo-ionized
plasma ($T \sim 2\times 10^4$ K) of a precursor of a CR-modified
shock \citep{Draine_McKee_93,ddk96}. In a cold partially ionized
plasma the generalized Ohm's law \citep[see,
e.g.,][]{Braginskii65,cowling76}
\begin{equation}\label{18}
\mathbf{E}+\frac{1}{c}\mathbf{u}\times\mathbf{B}
=\frac{\mathbf{j}}{\sigma}+\frac{1}{n_iec}
\mathbf{j}\times\mathbf{B}+\frac{F^2\tau_{ia}}{n_im_ic^2}\mathbf{B}
\times(\mathbf{j}\times\mathbf{B}).
\end{equation}
results in enhanced effective magnetic diffusion $\nu_{ef}$ in the
induction equation
\begin{equation}\label{28}
\nu_{ef}=\frac{c^2}{4\pi\sigma_\perp^{ef}},
\end{equation}
where
\begin{equation}\label{20}
\frac{1}{\sigma_\perp^{ef}}=\frac{1}{\sigma}+\frac{F^2B^2}{n_im_ic^2}.
\end{equation}
Here $F$ is the mass fraction of the neutrals, $\tau_{ia}$ is the
mean time between the ion-neutral collisions. The neutrals in cold
magnetized plasma of supernova shock precursors may strongly affect
both the growth and the dissipation of CR-driven waves \citep[see,
e.g.,][]{bt05,marcowithea06,revilleea07}. In a precursor of a
quasi-parallel shock the CR-driven turbulence is basically
incompressible and the energy dissipation rate $\Gamma^{in}$ in the
low $\beta$ limit can be estimated as
$$
\Gamma^{in} = \sigma_\perp^{ef}\,k_{\parallel}^2,
$$
while the ion-neutral dissipation of the compressible fluctuations
that are associated with long-wavelength CR-driven oblique wave
instabilities \citep[see, e.g.,][]{boe11,schureea12} are determined by
$$
\Gamma^{in} = \sigma_\perp^{ef}\,k^2.
$$

The magnetic turbulence dissipation term in Eq.~(\ref{eq_turb_evol})
due to effective magnetic diffusivity (Joule dissipation) is
determined by
\begin{equation}\label{28a}
L^{in}_{\rm turb}(x,\mathbf{k}) = \nu_{ef} k^2 W(x,\mathbf{k},t).
\end{equation}
To characterize a fraction of the magnetic energy dissipated in the
precursor of a CR-modified shock it is instructive to estimate the
minimal wavenumber $k_{m}$ of the fluctuations to be strongly
dissipated by ion-neutral friction while advecting through the
CR-shock precursor of the scale length $L_{CR} \sim c/v_{\rm
sh}\,\lambda_{\ast}$ as it is illustrated in Figure~\ref{sketch}, where
\begin{equation}\label{29}
 \nu_{ef}\,k_m^2 \times \frac{L_{CR}}{\rm v_{sh}}= 1.
\end{equation}
Then the maximal wavelength to be damped is
\begin{equation}\label{30}
\lambda_m = 2\pi/k_m \approx 2\pi\,F\, {\cal M}_{\rm a}^{-1}\,\times
\sqrt{c\tau_{ia}\lambda_{\ast}}.
\end{equation}
For a strong supernova shock of ${\cal M}_{\rm a} \approx$ 100  and
$F \approx$ 0.1 one may get $\lambda_m \lsim 10^{15}$ cm. Note,
however, that the length
 $\lambda_m$ is comparable to the ion-neutral collision length that
 is $\sim [F\,n_i\,\sigma_{in}]^{-1}$,  if the ion number density in the precursor
 $n_i \lsim 1\cmc$.
 Depending on the spectral shape of the CR-driven magnetic turbulence
in the shock precursor, the dissipated fraction ranges from a few
percent for a Kolmogorov-type spectrum to about 10\% for the
flatter spectra of strong turbulence. We neglected here the MHD wave
dissipation due to the thermal conduction and viscosity assuming a
cold plasma case \citep[see, e.g.,][for a thorough
discussion]{Braginskii65}. The magnetic power dissipated by the
ion-neutral collisions in the low $\beta$ plasma mainly heats ions.
The collisional damping discussed above results in a true irreversible
conversion of the magnetic turbulence free energy into the thermal
plasma energy. To the contrary,
the collisionless turbulence
dissipation processes  that we are going to discuss below can heat
electrons, thus being not completely irreversible. The collisionless
heating process just increases the
energy of quasi-thermal plasma components, but in general some
collisionality \citep[e.g.,][]{ramos11} is required to increase the
entropy and to reach the equilibrium distributions of plasma species.

\subsection{Collisionless heating of ions and electrons by magnetic turbulence}
\label{sec:heating} The CR-driven turbulence source
$S(x,\mathbf{k})$ in the shock precursor is expected to be
anisotropic in $\mathbf{k}$. The fast CR-current driven instability of
\citet{bell04} as well as the long-wavelength instability of
\citet{boe11} in the Bohm diffusion regime are mainly amplifying the
modes  with the wavevectors along the local magnetic field. The
acoustic instability of \citet{df86} driven by the CR-pressure gradient
is also anisotropic. The strong anisotropy of the magnetic
turbulence is observed in the solar wind where the outward flux
significantly exceeds the ingoing one.

The solar wind is one of the best laboratories to study anisotropic
magnetic turbulent dissipation and collisionless plasma heating
\citep[see,
e.g.,][]{leamonea98,Sahraouiea09,petrosyanea10,alexandrovaea11}.
Recently, \citet{Sahraouiea09} reported {\sl Cluster}
spacecraft measurement providing direct evidence that the
dissipation range of magnetofluid turbulence in the solar wind
extends down to the electron scales. Namely, they found two distinct
breakpoints in the magnetic spectrum at frequencies $f_p$ = 0.4
Hz and $f_e$ = 35 Hz, which correspond, respectively, to the
Doppler-shifted proton and electron gyroscales. Below $f_p$, the
spectrum follows a Kolmogorov-type scaling of power-law index about
-1.62. For  $f_p <f< f_e$ a second inertial range with a scaling
index -2.3 was established. Above $f_e$ the spectrum has a steeper
power law -4.1 down to the noise level of the {\sl Cluster}
detectors. The authors advocated a good agreement of the results
with theoretical predictions of a quasi-two-dimensional cascade into
Kinetic \alf Waves (KAW). \citet{chenea12} presented a measurement
of the scale-dependent, three-dimensional structure of the magnetic
field fluctuations in the inertial range of the solar wind
turbulence. The \alf-type fluctuations are three-dimensionally
anisotropic, with the sense
of this anisotropy varying from large to
small scales. At the outer scale, the magnetic field correlations
are longest in the local fluctuation direction. At the proton
gyroscale, they are longest along the local mean field direction and
shortest in the direction perpendicular to the local mean field and
the local field fluctuation. The compressive fluctuations are highly
elongated along the local mean field direction, although axially
symmetric in the perpendicular direction. Their large anisotropy may explain
why they are not heavily damped in the solar wind by the Landau damping.

Anisotropic wavenumber spectra that are broader in wavenumber
perpendicular to the background magnetic field than in the parallel one are
expected even in the symmetric MHD case studied by \citet{gs95}
where the oppositely directed waves carry equal energy fluxes. The
main features of the the symmetric (but anisotropic) MHD
turbulence model by \citet{gs95} are:

(i) a critical balance between the linear wave mode periods and
their nonlinear turnover timescales in the inertial-interval energy
spectrum;

(ii) the 'eddies' are elongated in the direction of the field on
small spatial scales with the scaling $k_\parallel \propto k_0^{1/3}
k_\perp^{2/3}$, where $k_0$ is the wavenumber corresponding to the
outer scale of the turbulence.

The three-dimensional simulations by \citet{choea04} revealed the
basic features of the \citet{gs95} model in the electron
magnetohydrodynamic turbulence.  The kinetic \alf wave and whistler
fluctuations are likely to make an important contribution to the
turbulence below the proton gyroscale \citep[see,
e.g.,][]{garyea12,sg12,boldyrevea12,rudakov12}. Particle-in-cell
simulations show that the anisotropic whistler turbulence heats the
electrons in the parallel direction as predicted by the linear theory
and that in the low $\beta$ plasmas the magnetic wavenumber spectrum
becomes strongly anisotropic with spectral index in the
perpendicular direction close to -4 \citep[see,
e.g.,][]{garyea12,sg12}. Microscopic two-dimensional particle-in-cell
simulations  of whistler turbulence were carried out by \citet{sg12}
in a collisionless homogeneous magnetized plasma to study the
electron and ion heating dependence on the plasma magnetization
parameter $\beta$. They demonstrated that at higher values of $\beta$
the magnetic energy cascade in the perpendicular direction becomes
weaker and leads to more isotropic wavenumber spectra. The electron
energy ratio between parallel and perpendicular components becomes
closer to unity at higher $\beta$. Three-dimensional  particle-in-cell
plasma simulations of decay of initial long wavelength whistler
fluctuations into a broadband, anisotropic, turbulent spectrum at
shorter wavelengths via a forward cascade were performed by
\citet[][]{garyea12}. The simulations demonstrated a picture
qualitatively similar to that in 2D but somewhat stronger anisotropy
of the resulting 3D turbulence comparing to the similar 2D runs.
They showed a clear break in the perpendicular wavenumber spectra
qualitatively similar to that measured in the electron scale
fluctuations in the solar wind. Earlier \citet[]{quataert98} and \citet{qg99}
discussed the beta-dependence of particle heating by turbulence in
advection-dominated accretion flows. They found that for $\beta
\sim$ 1, i.e. approximately equipartition magnetic fields, the
turbulence primarily heats the electrons. For weaker magnetic
fields, the protons are primarily heated. The division between
electron and proton heating occurs between $5 < \beta <100$,
depending on unknown details of how \alf waves are converted into
whistlers at the proton gyroscales.

The cascade of \alf waves, which are weakly damped down to the
scale of the proton gyro-radius $k_\perp \rho_i \sim 1$ is a subject
of gyro-kinetic models \citep[see for a review][]{gyrokin09}. The
models can be of interest for the parallel shock precursor heating
assuming an efficient cascading of the CR-driven magnetic
fluctuations down to the $k_\perp \rho_i \sim 1$ regime. The
cascading still remains to be demonstrated since the CR-current driven
modes are very different from the standard \alf waves. If the
cascading occurs, then the continuity equation
(\ref{eq_turb_evol}) can be reduced to the equation for $
b_k^2(k_\perp)= k_\perp^2 \int dk_\parallel W(\mathbf{k})$ -- the
energy density of the anisotropic magnetic turbulence as a function of the
perpendicular wavenumber \citep[see, e.g.,][]{howes10,cvb12}:
\begin{equation}
\frac{\partial b_k^2}{\partial t} + k_\perp \frac{\partial
\epsilon(k_\perp) }{\partial k_\perp} = S(k_\perp) -
{\Gamma(k_\perp)} b_k^2, \label{kin_perp}
\end{equation}
where the energy injection rate is $S$ (non-zero only at the driving
scale $k_\perp=k_0$), the linear energy damping rate is
$\Gamma$, and $\epsilon(k_\perp)$ is the energy cascade rate. To
specify the energy cascade rate both advection and diffusion in the
wavenumber space models are used (see, e.g., \citet{cvb12} for a
recent discussion). Assuming critical balance at all scales and
using the energy cascade rate in the form of the advection in the
wavenumber space:
$$
\epsilon(k_\perp) = C_1^{-3/2} k_\perp \overline{\omega} b_k^3
$$
\citet{howes10} obtained the steady state solution for the energy
cascade rate as
\begin{equation}
\epsilon(k_\perp) = \epsilon_0
\exp{\left\{-\int_{k_0}^{k_\perp}C_1^{3/2}C_2
\frac{\overline{\Gamma}(k_\perp ')}{\overline{\omega}(k_\perp ')}
\frac{dk_\perp '}{k_\perp '}\right\}}, \label{howes3}
\end{equation}
where $C_1$ and $C_2$ are the dimensionless Kolmogorov constants
($C_1=1.96$ and $C_2=1.09$) and $\epsilon_0$ is the rate of energy
input at $k_0$. \citet{howes10} used the normalized eigenfrequencies
$\overline{\omega}(k_\perp)$ from the linear gyrokinetic dispersion
relations and the damping rates $\overline{\Gamma}_s$ due to
different plasma species from \citet[][]{howesea06} and
Eq.~(\ref{howes3}) (where $s=i,e$), to calculate the spectrum of
heating by species 
$$
Q_s(k_\perp)=C_1^{3/2}C_2(\overline{\Gamma}_s/\overline{\omega}) 
\epsilon(k_\perp)/k_\perp.
$$

The ion damping peaks at $k_\perp \rho_i \sim 1$, while the electron
damping peaks at $k_\perp \rho_i \gg 1$ unless $T_i/T_e \lsim
m_e/m_i$. The energy that passes through the peak of the ion damping
at $k_\perp \rho_i \sim 1$ would lead to electron heating assuming
both the cascading and the damping times are shorter than the
advection time through the shock precursor. Then the total
(integrated over $k_\perp\, \rho_i \gsim 1$) ion-to-electron heating
rate  due to the kinetic dissipation of the turbulent cascade,
$Q_i/Q_e(\beta_i, T_i/T_e)$ can be approximately fitted with
\begin{equation}
Q_i/Q_e = c_1\frac{c_2^2 + \beta_i^p}{c_3^2 + \beta_i^p}
\sqrt{\frac{m_i T_i}{m_e T_e}} e^{-1/\beta_i}, \label{eq:fit}
\end{equation}
where $c_1=0.92$, $c_2=1.6/(T_i/T_e)$, $c_3=18+5 \log (T_i/T_e)$,
and $p=2-0.2 \log (T_i/T_e)$.  A slightly better fit for $T_i/T_e
<1$ occurs with the coefficients $c_2=1.2/(T_i/T_e)$ and $c_3=18$
\citep{howes10}. The model is valid for the parameter range $0.01
\le \beta_i \le 100$ and $0.2 \le T_i/T_e \le 100$. The heating rate
$Q_i/Q_e$ is an approximately monotonic function of $\beta_i$ that
is only weakly dependent of $T_i/T_e$.

The simplified model of the electron and ion heating discussed above
assumed the nonlinear collisionless cascading from the energy
containing scale of wavenumber $k_\ast R_g \sim 1$ where the
amplitude of amplified magnetic field is $\delta B_\ast$ that is
determined by the gyroscale $R_g$ of the energy containing
accelerated particles to the thermal ion gyroscale $k_\perp \rho_i
\sim 1$. In the shock frame the advection time of the incoming
plasma through the CR precursor $\tau_{\rm adv} \sim
\lambda_{\ast}\,c/v_{\rm sh}^2$. Therefore for efficient heating of
the plasma species the cascading time $\tau_{c}$ must be shorter
than $\tau_{\rm adv}$. If the cascading time is determined by
the turn-over time of the energy containing magnetic "vortex"
$\tau_{c}^{-1} \sim k_\ast \delta B_\ast/\sqrt{4\pi\rho}$ then the
condition of efficient plasma heating in the shock precursor by the
CR-driven turbulence can be written as
\begin{equation}
\frac{\tau_{\rm adv}}{\tau_{c}} =  \eta\, \frac{c}{v_{\rm sh}^2}\,
\frac{\delta B_\ast}{\sqrt{4\pi\rho}} > 1, \label{cond1}
\end{equation}
where $\eta = \lambda_{\ast}/R_g >$ 1. The amplitude of amplified
magnetic field likely scales with the shock velocity as $\delta
B_\ast \propto v_{\rm sh}^{\rm b}$, where $1\leq {\rm b} \leq 1.5$
\citep[see, e.g.,][]{vink12}. The condition (\ref{cond1}) predicts
a less efficient plasma heating in the CR-precursor of the faster
shocks.

Heat conduction that we did not discuss here may play a role in the
electron temperature distribution in the shock precursor
\citep[c.f.][]{breechea09}. The nonlinear dynamics of the the CR-driven
magnetic fluctuations in the shock precursor deserves thorough modeling.
\citet[][]{msd12} have obtained fully nonlinear exact solutions  of the
{\sl ideal} 1D-MHD supported by the CR return current. The solutions occur
as localized spikes of circularly polarized \alf envelopes (solitons or
breathers). The sufficiently strong solitons in the model run ahead of the
main shock and stand in the precursor, being supported by the return
current. The CR-shock precursor in the model is dissipationless.

The electron and ion temperatures in the shock precursor determine
the injection of particles into the CR acceleration regime. The
temperatures can be tested by optical spectroscopy of supernova
shocks. CR-modification of shocks with modest speed of a few hundred
$\kms$ may yield lower post-shock temperatures and thus make the
post-shock flow switch to a radiative regime.

\section[Spectroscopy of a CR-modified radiative shock]{Spectroscopy of a CR-modified radiative shock
\footnote{ This section uses some results of numerical code
\texttt{SHELLS} currently developed by A.M.~Bykov,  A.E.~Vladimirov,
and A.M.~Krassilchtchikov. This code is used to model the
steady-state structure and broadband continuum and line emission
spectra of radiative shocks. The model accounts for the impact of
efficient CR acceleration on the shock compression ratio and the
postshock flow. The code incorporates modern atomic data and allows us
to consider in detail the non-equilibrium microscopic,
thermodynamical and radiative processes that determine the plasma
flow.} } \label{sec:1}

Consider a one-dimensional flow around a collisionless shock
consisting of three zones: a) the pre-shock, where the unperturbed
interstellar matter is preionized and preheated by the radiation
(and energetic particles) generated in the downstream and where
strong fluctuations of magnetic field may be generated by the CR
anisotropy, b) the thin shock front (a "viscous jump") where a
substantial part of the kinetic energy of the bulk upstream flow is
converted into energy of thermal motions, and c) the post-shock,
where the hot flow cools down, radiating continuum and line
emission. We discuss here a class of shock flows in partly ionized
media, the so-called {\it radiative shocks}, where the power
radiated away from the post-shock flow is a sizeable fraction of the
total kinetic and magnetic power dissipated at the shock \citep[see,
e.g.,][]{spitzer78,Draine_McKee_93}.

\begin{figure*}
\includegraphics[width=0.95\textwidth]{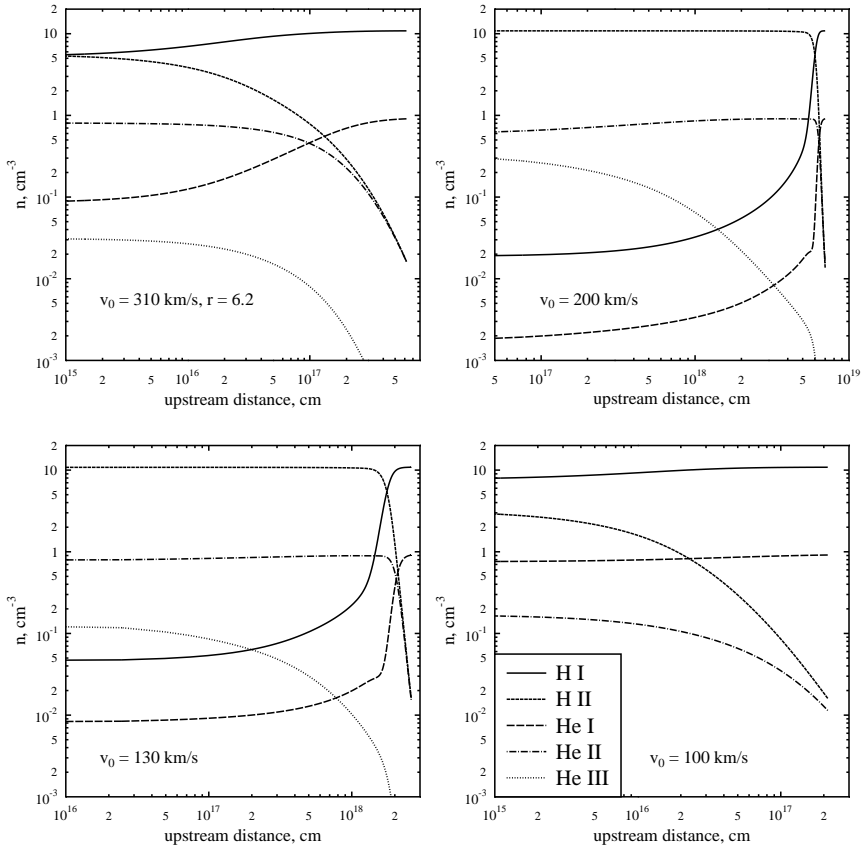}
\caption{Hydrogen and helium pre-shock (photo)ionization in the
shock upstream for four sets of shock parameters simulated with the
model of radiative shock accounting for the effect of CR escape on
the shock compression $r$ (the CR-modified case is shown in the top
left panel). Within the presented model $r$ = 6.2 corresponds to
30\% of the flow energy being converted into CRs.}
\label{fig:ion}       
\end{figure*}

Without a significant impact on accuracy, one may assume that the
pre-shock is isothermal, and the flow is in a steady state
(see also a comment about the equilibrium between the pre-shock ionization state and the ionizing flux
at the end of Section~\ref{sec: photo}) . However,
the ionization state of the plasma is non-uniform: the ionization
level increases toward the subshock, as the gas absorbs the ionizing
radiation flux emerging from the hot downstream region. This
photoionization does not have a significant impact on the thermal
state of the gas, because the hot electrons produced by
photoionization do not have sufficient time to collisionally
equilibrate with the atoms and ions. CR particles may contribute to
gas ionization and heating in the interstellar clouds and shells
\citep[e.g.,][]{spitzer78,bb94}. The effect of CRs may be important
in the vicinity of fast Balmer-type interstellar shocks as it has
been demonstrated by \citet[][]{morlinoea12}, who did not account
for the photo processes, though. Radiative shocks are expected to
occur in relatively dense environment and, therefore, have
velocities typically well below 1,000 $\kms$. In such an
environment, photo processes can provide a high ionization degree
of the upstream gas for sufficiently high shock speeds.
If CR acceleration is weak, the photoionization becomes significant
for shock velocities above 100 $\kms$.
However, if CR acceleration is efficient, the downstream temperature is reduced and the shock compression is increased,
which means that strong photoionization in the precursor occurs at much greater shock speeds
See Fig.~\ref{fig:ion} for an illustration of the effect of CR acceleration on precursor ionization. We
demonstrate below the effect of the CR fluid on the shock
compression and spectra of radiative shocks.

The jump conditions at the shock are either given by the
Rankine-Hugoniot equations, or, if the shock structure is assumed to
be modified by accelerated particles \citep[e.g.,][ and references
therein]{Bykov_04,vbe08}, the compression ratio can be parameterized
by the fraction $Q_{\rm esc}$ of bulk flow energy carried away by these
particles. The subshock is the standard gas viscous shock of a Mach
number ${\cal M}_{\rm sub}$. For that simplified {\sl two-fluid}
model of a strong CR-modified shock the effective ion temperature in
the downstream $T^{(2)}_{\rm i}$ can be estimated for a shock of a
given velocity, if $r_{\rm tot}$ and $r_{\rm sub}$ are known:
\begin{equation}
T^{(2)}_{\rm i} \approx \phi({\cal M}_{\rm sub})  \cdot \frac{ \mu~
v_{\rm sh}^2}{\gamma_{\rm g}r_{\rm tot}^2(v_{\rm sh})} ,~~{\rm
where}~~ \phi({\cal M}_{\rm sub}) = \frac{2 \gamma_{\rm g} {\cal
M}_{\rm sub}^2 - (\gamma_{\rm g} -1)}{(\gamma_{\rm g} -1){\cal
M}_{\rm sub}^2 + 2}. \label{eq:tcr}
\end{equation}

Single fluid strong shock heating represents the limit ${\cal
M}_{\rm sub} = {\cal M}_{\rm s} \gg$ 1, since there is no precursor
in that case the temperature behind a strong shock is determined by
the standard scaling
\begin{equation}
 T^{(2)} \approx 2\cdot \frac{(\gamma_{\rm g}
-1)}{(\gamma_{\rm g} +1)^2}~\mu v_{\rm sh}^2 = 1.38 \cdot
10^7~v^2_{s8}~(K), \label{eq:rht1}
\end{equation}

In single-fluid systems the compression ratio $r_{\rm tot} = r_{\rm
sub} \rightarrow (\gamma_{\rm g} +1)/(\gamma_{\rm g} -1)$ does not
depend on the shock velocity and Eq.(\ref{eq:tcr}) reduces to
Eq.(\ref{eq:rht1}). However, in multi-fluid shocks the total
compression ratio depends on the shock velocity and could be
substantially higher than that in the single-fluid case.
Consequently, the post-shock temperature in a multi-fluid shock is
lower than the post-shock temperature in a single-fluid shock of the
same speed. This allows us to determine the CR acceleration efficiency
using observations of post-shock temperatures and shock speeds, or
the entropy profiles in the accretion shocks of clusters of galaxies
\citep[][]{bykov_05,bdd08,bbrr12,foy13}. It is convenient to
introduce the scaling $r_{\rm tot}(v_{\rm sh}) \propto v_{\rm
sh}^{\xi}$ to describe the different cases of strong shock heating
\citep{bdd08}. Then from Eq.~(\ref{eq:tcr}) $T^{(2)}_{\rm i} \propto
\phi({\cal M}_{\rm sub})\cdot v_{\rm sh}^{2(1-\xi)}$. The subshock
Mach number ${\cal M}_{\rm sub}$ depends, in general, on ${\cal
M}_{\rm s}$ and ${\cal M}_{\rm a}$. Thus, the index $\sigma$
approximates the velocity dependence of $\phi({\cal M}_{\rm sub})
\propto v_{\rm sh}^{\sigma}$. Finally, if $T^{(2)}_{\rm i} \propto
v_{\rm sh}^a$, then the index $a = 2(1-\xi) + \sigma$ . For the case
of shock precursor heating by CR generated \alf waves, the index $a
\approx$ 1.25 \citep{bykov_05}. The effects of neutrals due to
charge exchange in the shock downstream with heated ions that
results in a flux of high-velocity neutrals that return upstream
were studied by \citet{blasiea12} and \citet{ohira12}. They found
that the return flux of neutrals may result in the reduced shock
compression ratio and spectral steepening of test particles
accelerated at the shocks slower than about 3000 $\kms$. The return
flux of neutral atoms may also affect the radiation spectrum of the
post-shock flow that we are modeling.

The ratio of the electron to ion temperature immediately after the subshock, $T_e/T_i=\delta_e$,
is considered as a free parameter in our model. It can be estimated from observations as discussed in Section~\ref{sec:interaction}.
It is varied in the range from $\sqrt{m_e/m_p} \approx 0.023$ to 1.
Observations typically show low values of $\delta_e$ in shocks
faster than 1000 $\rm km~s^{-1}$, and values closer to 1 in slower
shocks \citep[e.g.,][]{Ghavamian_ea_01, Ghavamian_ea_07,
Rakowski_ea_08, HKV_10}.

The post-shock plasma is treated as a stationary two-temperature
single fluid flow consisting of ions, electrons, and neutral atoms
that are ideal nonrelativistic gases. The neutral and ion
temperatures may actually differ \citep{Heng_ea_07,
vanAdelsberg_ea_08}, but that is only important just behind a very
fast shock in partially neutral gas, and it matters mainly for
diagnostics based on the H$\alpha$ line profile. For the considered
here ranges of shock speeds and matter densities the temperature
equilibration scales are below 10$^{15}$ cm$^{-2}$, while the line
emission zone typically lies well above 10$^{16}$ cm$^{-2}$
downstream.

Let $z$ and $\mu$ be the average charge and mass of an ion,
$\zeta \equiv m_e / \mu$ -- the average electron to ion mass ratio. Then
\be
n_e =  z n_i,\ \rho_e = m_e n_e = z m_e n_i, \\ \nonumber
\rho_i = \mu n_i, \\ \nonumber
\rho = n_i (z m_e + \mu) = n_i \mu (z \zeta +1) \equiv n_i \mu / M(z), \\ \nonumber
M(z) = 1/(z \zeta +1).
\ee
Let $\rho_0, v_0$, and $B_0$ be the values of density, velocity, and
frozen-in transverse magnetic field just before the shock, and
$v_a, T_i^a$ be the values of flow velocity and ion temperature
in the immediate post-shock defined by jump conditions.

Then, the flow evolution downstream can be described
by the following system of equations.

\be
\rho v = {\rm const} = \rho_0 v_0, \label{eq1} \\
\rho v^2 + p_e + p_i + p_m + p_{CR} = {\rm const} \equiv \Pi \rho_0 v_0^2, \label{eq2} \\
{3\over 2} n_i v {dT_i \over dx} = - n_i T_i {dv \over dx} - {3 m_e \over \mu}{n_e \over \tau_{ei}} (T_i-T_e),
                    \label{eq3}\\
{3\over 2}n_e v {dT_e \over dx} = - n_e T_e {dv \over dx} + {3m_e \over \mu}{n_e \over \tau_{ei}} (T_i-T_e)
      - \Lambda - {3\over 2} n_i T_e v {dz \over dx}, \label{eq4}
\ee
where
$ p_e + p_i = n_i k_B (z T_e + T_i) = \rho k_B M (z T_e + T_i) /\mu$,
the magnetic pressure $ p_m = B_0^2 \rho^2 / ( 8 \pi \rho_0^2 ) $,
$\Pi = v_a / v_0 + B_0^2 / ( 8 \pi \rho_0 v_0 v_a ) +
 k_B M T_i^a ( z \delta_e + 1) / ( v_0 v_a \mu ) + Q_{\rm esc}/2 $,
$p_{CR} = (Q_{\rm esc}/2)\cdot \rho_0 v_0^2 (v_a / v )^{4/3}$,
$\Lambda$ is the cooling function,
$\tau_{ei}$ is the electron-ion equilibration time,
and the last term of equation (\ref{eq4}) denotes electron
cooling due to ionization.

The cooling term $\Lambda$ is calculated as
\be
\Lambda = n^2 \Lambda_{coll} + n_e \sum_{i,j} ( n_{i,j} \alpha_{i,j} E_{i,j}^{rec}
          + n_{i,j} C_{i,j} h \nu_j ) - \\ \nonumber
          - 4 \pi \sum_{i,j} \int_{\nu_j}^{\infty} d\nu \cdot \sigma_{i,j}^{ph}
        ( 1 - \nu_j / \nu ) n_{i,j} J_{\nu},
\ee
where $i$ denotes the chemical element and $j$ denotes the ionization state
of an ion ($j$=0 corresponds to a neutral atom).
Here $\Lambda_{coll}$ is due to electron-ion collisions including
electron bremsstrahlung and line emission of the ions excited by electron impact,
though it does not include emission due to radiative recombination;
$\alpha_{i,j}$ is the recombination rate $\{j+1\} \rightarrow \{j\}$;
$E_{i,j}^{rec}$ is the average energy on the recombining electrons;
$C_{i,j}$ is the rate of collisional ionization;
$\sigma_{i,j}^{ph}$ is the photoionization crossection of the ion state $j$;
$J_{\nu}$ is the angle-averaged density of ionizing radiation
at frequency $\nu$:
\be
J_{\nu}(x) = {1\over 4\pi}\ \int^1_{-1} 2\pi\ I_{\nu}(\mu,x)\ d\mu.
\ee
where $\mu = cos(\theta)$, $\theta$ is the angle between the normal
to the shock front and the direction of emitted photons.

The evolution of the ionization state $j$ of an ion $i$ in the downstream
flow is determined as
\be
v {d n_{i,j} \over dx} = n_e \left ( n_{i,j-1} \tilde C_{i,j-1} -
       n_{i,j} \tilde C_{i,j} - n_{i,j} \tilde \alpha_{i,j} +
        n_{i,j+1} \tilde \alpha_{i,j+1} \right) + \\ \nonumber
       + n_{i,j-1} R_{i,j-1} - n_{i,j} R_{i,j} + \\ \nonumber
     + \sum_{s = H,He,He^{+} } n_s \left[ n_{i,j-1} V^{ion}_{s,i,j-1} -
       n_{i,j} ( V^{ion}_{s,i,j} + V^{rec}_{s,i,j} )
       + n_{i,j+1} V^{rec}_{s,i,j+1} \right],
\ee
where $ R_{i,j} $ is the photoionization rate, $V^{ion}_s$ and $V^{ion}_s$
are the rates of ionization and recombination via charge exchange reactions
with the ion $s$, $\tilde C_{i,j} = C_{i,j} +C_{i,j}^{\rm auto}$, where
$C_{i,j}^{\rm auto}$
is the autoionization rate, $\tilde \alpha_{i,j} = \alpha_{i,j} + \alpha_{i,j}^{2e}$,
where $ \alpha_{i,j}^{2e} $ is the dielectronic recombination rate.

To obtain the ionizing radiation field a transfer equation can be
solved both in the downstream and in the upstream: \be \cos{\theta}
\cdot {d I_\nu \over dx} = j_\nu - \kappa_\nu I_\nu, \ee where
$\theta$ is the angle between the normal to the shock front and the
direction of emitted photons. In this equation, the absorption coefficient $\kappa_\nu =
\sum_{i,j} n_{i,j} \sigma^{ph}_{i,j}$ is determined by bound-free transitions
in all ion species. The ionizing emission
is generated as a) permitted ultraviolet and optical
lines excited by an electronic impact, b) recombination line
cascades of hydrogen and helium, c) free-free continuum of electrons
scattering at ions, d) 2-photon continuum emission of H- and He-like
ions where metastable levels are collisionaly populated, e)
recombination continuum emission.

\begin{figure*}
\includegraphics[width=0.95\textwidth]{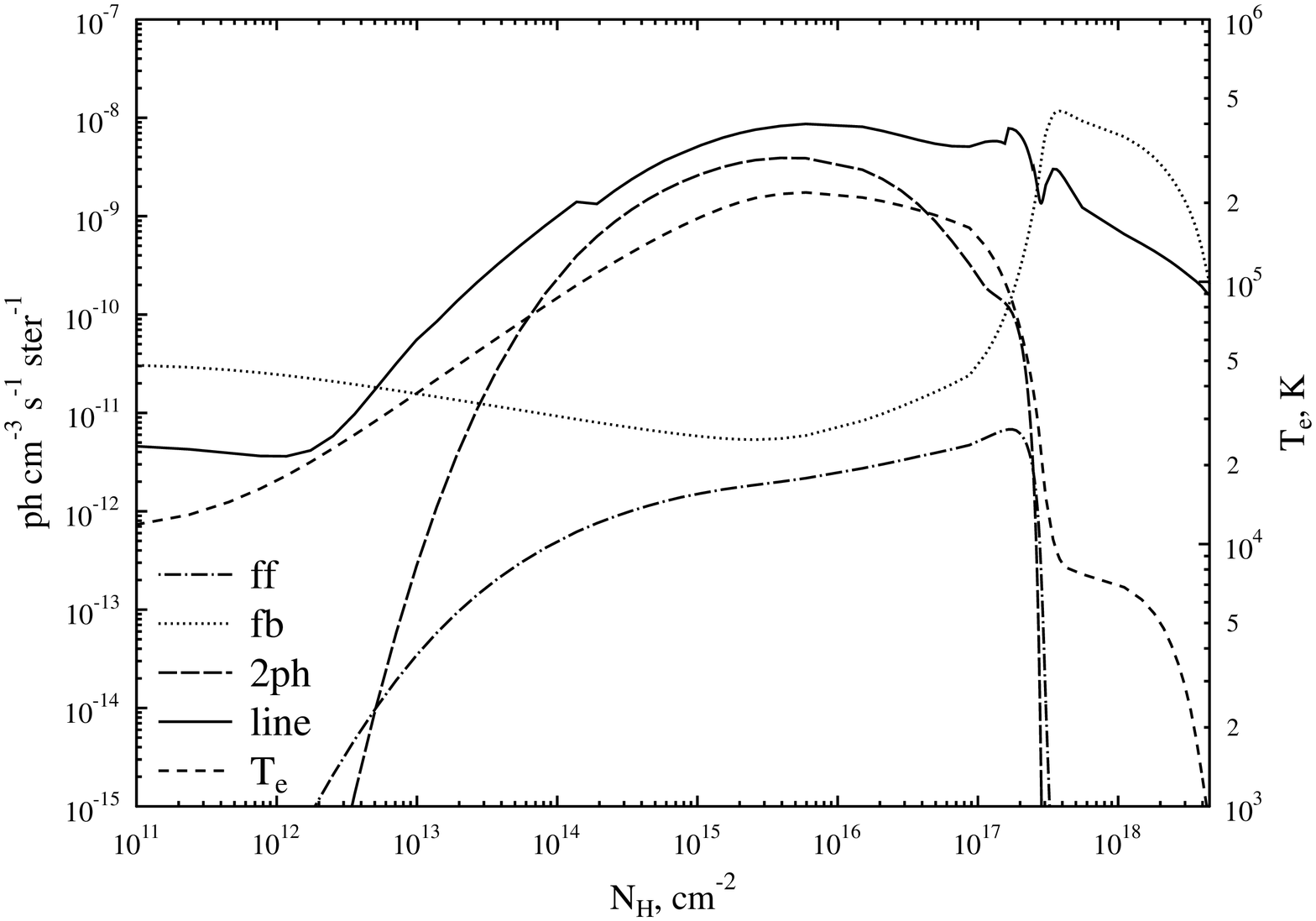}
\caption{Production rate of ionizing photons ($h\nu >$ 1~Ry) in the
downstream of a 130 km/s shock penetrating into a 10 cm$^{-3}$
medium of solar abundance. Here "fb" denotes recombination continuum
emission, "ff" denotes free-free continuum of electrons scattering
at ions, "2ph" stands for 2-photon continuum emission of H- and
He-like ions, and "line" denotes permitted ultraviolet lines excited
by an electronic impact. The dashed line illustrates the simulated
electron temperature profile in the downstream. The photons produced
in the optically thin part of the downstream are ionizing the
upstream flow.}
\label{fig:ioniz}       
\end{figure*}

The relative contribution of each of the mechanisms to the ionizing
photon field is illustrated in Figure~\ref{fig:ioniz}. The
dominating 2-photon continuum emission comes mainly from He~I and
He~II and the dominating ultraviolet line emission comes mainly from
He~I, He~II, and oxygen ions upto O~V. It should be noted that in
the presented model the electron temperatures just after the shock
are assumed to be low ($\delta_e \sim$ 10$^{-2}$) and equilibrate
with the ion temperatures downstream via Coulomb collisions.

\begin{figure*}
\includegraphics[width=0.95\textwidth]{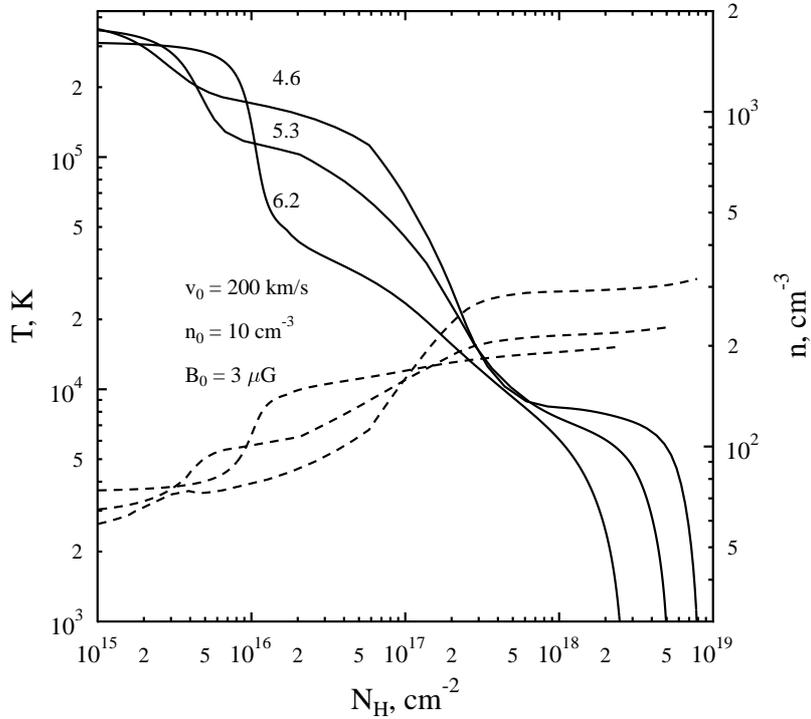}
\caption{ Gas temperature (solid lines) and density (dotted lines)
in the downstream cooling region of three 200 km/s shocks. Three
lines correspond to three models with different assumptions
regarding the efficiency of CR acceleration. Within the presented
model $r$ = 6.2 corresponds to $Q_{\rm esc}$ = 30\% of the flow
energy being converted into CRs, $r$ = 5.3 corresponds to $Q_{\rm
esc}$~=~20\%, and $r$ = 4.6 corresponds to $Q_{\rm esc}$ = 10\%. }
\label{fig:post-shock}       
\end{figure*}

\begin{table}
\begin{tabular}{lrrrrlrrrr}

line        & $\lambda$   & r = 4.6  &  r = 5.3 & r = 6.2 & line
& $\lambda$   & r = 4.6  &  r = 5.3 & r = 6.2 \\ \hline
%
He II       & 304 \AA   & 2640\dd  &  750\dd & 1.1   & $[$S II$]$  & 6718 \AA  &   67.1   &   27.8  & 16.1 \\  
O IV        & 554 \AA   &  201\dd  &   14.7  & $\cdots$  & $[$S II$]$  & 6733 \AA  &   43.9   &   18.9  & 10.6 \\ 
He I        & 584 \AA   &   72.5   &  212\dd  & 103\dd  & $[$Fe II$]$ & 1.28 \mm  &  126\dd  &   85.1 & 51.1 \\ 
O V         & 630 \AA   &   24.5   & $\cdots$ & $\cdots$ & $[$Fe II$]$ & 1.62 \mm  &   29.0   &   19.4 & 11.6  \\ 
O III       & 833 \AA   &   98.4   &  142\dd &  0.5  & $[$Fe II$]$ & 5.30 \mm  &   95.2   &   78.2  & 49.4 \\ 
C III       & 977 \AA   &  637\dd  &  770\dd & 52.1 & $[$S IV$]$  & 10.51 \mm &    2.1   &    2.0  & 0.1 \\ 
O VI        & 1034 \AA  & $\cdots$ & $\cdots$ & $\cdots$ & $[$Ne II$]$ & 12.81 \mm &   40.3   &   26.6  & 8.1 \\ 
Ly$\alpha$  & 1216 \AA  & 5500\dd  &10000\dd & 17000 & $[$Ne III$]$& 15.55 \mm &   29.7   &   12.6  & $\cdots$ \\ 
N V         & 1240 \AA  &    2.9   &    0.1  &  $\cdots$ & $[$Fe II$]$ & 17.94 \mm &   10.1   &    8.1  & 5.1 \\ 
Si IV       & 1397 \AA  &  114\dd  &  122\dd & 2.1 & $[$S III$]$ & 18.71 \mm &    1.3   &    1.8  & 1.1 \\ 
C IV        & 1549 \AA  &  858\dd  &  190\dd & 0.1 & $[$Fe II$]$ & 24.52 \mm &    1.8   &    1.4  & 0.9 \\ 
He II       & 1640 \AA  &   58.5   &   12.6  & $\cdots$ & $[$O IV$]$  & 25.91 \mm &   12.4   &    2.9 & $\cdots$ \\ 
C III $]$   & 1908 \AA  &  219\dd  &  290\dd & 50.9 & $[$Fe II$]$ & 26.00 \mm &   47.5   &   43.5  & 29.6 \\ 
$[$O II$]$  & 3729 \AA  &  236\dd  &  221\dd & 289\dd & $[$S III$]$ & 33.50 \mm &    1.7   &    2.4  & 1.6 \\ 
$[$Ne III$]$& 3870 \AA  &   23.6   &   14.9  & $\cdots$ & $[$Si II$]$ & 34.82 \mm &  218\dd  &  122\dd & 51.8  \\ 
$[$Ne III$]$& 3969 \AA  &    7.1   &    4.5  & $\cdots$ & $[$Fe II$]$ & 35.35 \mm &   10.2   &    9.2  & 6.2 \\ 
He II       & 4687 \AA  &    6.2   &    1.1  & $\cdots$ & $[$Ne III$]$& 36.02 \mm &    2.6   &    1.1  & $\cdots$ \\ 
H$\beta$    & 4861 \AA  &  100\dd  &  100\dd & 100\dd  & $[$N III$]$ & 57.34 \mm &    1.7   &    1.6  & 0.1 \\ 
$[$O III$]$ & 4960 \AA  &   48.3   &   49.0  & 0.4 & $[$O I$]$   & 63.19 \mm &   28.7   &   12.7 & 4.9 \\ 
$[$O III$]$ & 5008 \AA  &  144\dd  &  146\dd & 1.0 & $[$O III$]$ & 88.36 \mm &    9.8   &    8.2  & 0.1 \\ 
$[$O I$]$   & 6300 \AA  &   27.4   &    7.0  & 3.1 & $[$N II$]$  & 121.8 \mm &    6.2   &    3.8  & 2.0 \\ 
$[$O I$]$   & 6363 \AA  &   27.3   &    6.9  & 3.1 & $[$O I$]$   & 145.5 \mm &    3.8   &    1.9 & 0.8 \\ 
H$\alpha$   & 6565 \AA  &  302\dd  &  310\dd & 321\dd & $[$C II$]$  & 157.7 \mm &   18.3   &   13.5  & 7.9 \\ 
$[$N II$]$  & 6550 \AA  &   46.6   &   24.6  & 19.9 & H$\beta$    & 4863 \AA  &    6.38   &   5.61 & 5.05 \\ 
$[$N II$]$  & 6585 \AA  &  142\dd  &   75.1  & 60.7 & & & & & \\ 
\end{tabular}

\caption{Model line strengths in percent of H$\beta$ flux for $v_0$
= 200 km/s at CR-modified compression ratios $r$ = 4.6, 5.3, 6.2.
The last line presents absolute flux in H$\beta$ 4861 \AA line in
10$^{-6}$ \ecss. \label{table-lines-rel}}
\end{table}

All the ionizing lines except Ly$\alpha$ were considered optically thin. The
optically thick case of Ly$\alpha$ was treated according to the
standard 2-level formalism adopted from \citet{Mihalas_84}.


\begin{table}
\begin{tabular}{llllll}

%
He II       & 304 \AA   & 2600\dd  & $[$N II$]$  & 6585 \AA  &  139\dd  \\ 
O IV        & 554 \AA   &  597\dd  & $[$S II$]$  & 6718 \AA  &   71.6  \\ 
He I        & 584 \AA   &   45.1      & $[$S II$]$  & 6733 \AA  &   45.1  \\ 
O V         & 630 \AA   &  509\dd  & $[$Fe II$]$ & 1.28 \mm  &  288\dd  \\ 
O III       & 833 \AA   &  179\dd   & $[$Fe II$]$ & 1.62 \mm  &   69.4  \\ 
C III       & 977 \AA   &  438\dd   & $[$Fe II$]$ & 5.30 \mm  &  204\dd \\ 
O VI        & 1034 \AA  &    0.6     & $[$S IV$]$  & 10.51 \mm &    2.3  \\ 
Ly$\alpha$  & 1216 \AA  &  641\dd & $[$Ne II$]$ & 12.81 \mm &   61.2 \\ 
N V         & 1240 \AA  &   14.3     & $[$Ne III$]$& 15.55 \mm &   24.4  \\ 
Si IV       & 1397 \AA  &  138\dd  & $[$Fe II$]$ & 17.94 \mm &   36.4  \\ 
C IV        & 1549 \AA  & 1080\dd & $[$S III$]$ & 18.71 \mm &    1.9  \\ 
He II       & 1640 \AA  &   74.9      & $[$Fe II$]$ & 24.52 \mm &    6.7  \\ 
C III $]$   & 1908 \AA  &  152\dd  & $[$O IV$]$  & 25.91 \mm &   15.0 \\ 
$[$O II$]$  & 3729 \AA  &  254\dd & $[$Fe II$]$ & 26.00 \mm &  114\dd \\ 
$[$Ne III$]$& 3870 \AA  &   23.6   & $[$S III$]$ & 33.50 \mm &    1.8  \\ 
$[$Ne III$]$& 3969 \AA  &    7.1    & $[$Si II$]$ & 34.82 \mm &  143\dd \\ 
He II       & 4687 \AA  &    8.9        & $[$Fe II$]$ & 35.35 \mm &   25.2  \\ 
H$\beta$    & 4861 \AA  &  100\dd  & $[$Ne III$]$& 36.02 \mm &    2.1  \\ 
$[$O III$]$ & 4960 \AA  &   52.2    & $[$N III$]$ & 57.34 \mm &    1.0  \\ 
$[$O III$]$ & 5008 \AA  &  156\dd & $[$O I$]$   & 63.19 \mm &   47.4  \\ 
$[$O I$]$   & 6300 \AA  &   35.7    & $[$O III$]$ & 88.36 \mm &    4.0  \\ 
$[$O I$]$   & 6363 \AA  &   35.5    & $[$N II$]$  & 121.8 \mm &    1.9  \\ 
H$\alpha$   & 6565 \AA  &  302\dd & $[$O I$]$   & 145.5 \mm &    5.6  \\ 
$[$N II$]$  & 6550 \AA  &   45.5    & $[$C II$]$  & 157.7 \mm &    7.3  \\ 
\end{tabular}
\caption{Model line intensities expressed in percent of the 4861 \AA\
H$\beta$ line instensity, which amounts here to
7.87$\times$10$^{-6}$ \ecss.\label{table-lines-abs}}
\end{table}

\begin{figure*}
\includegraphics[width=0.95\textwidth]{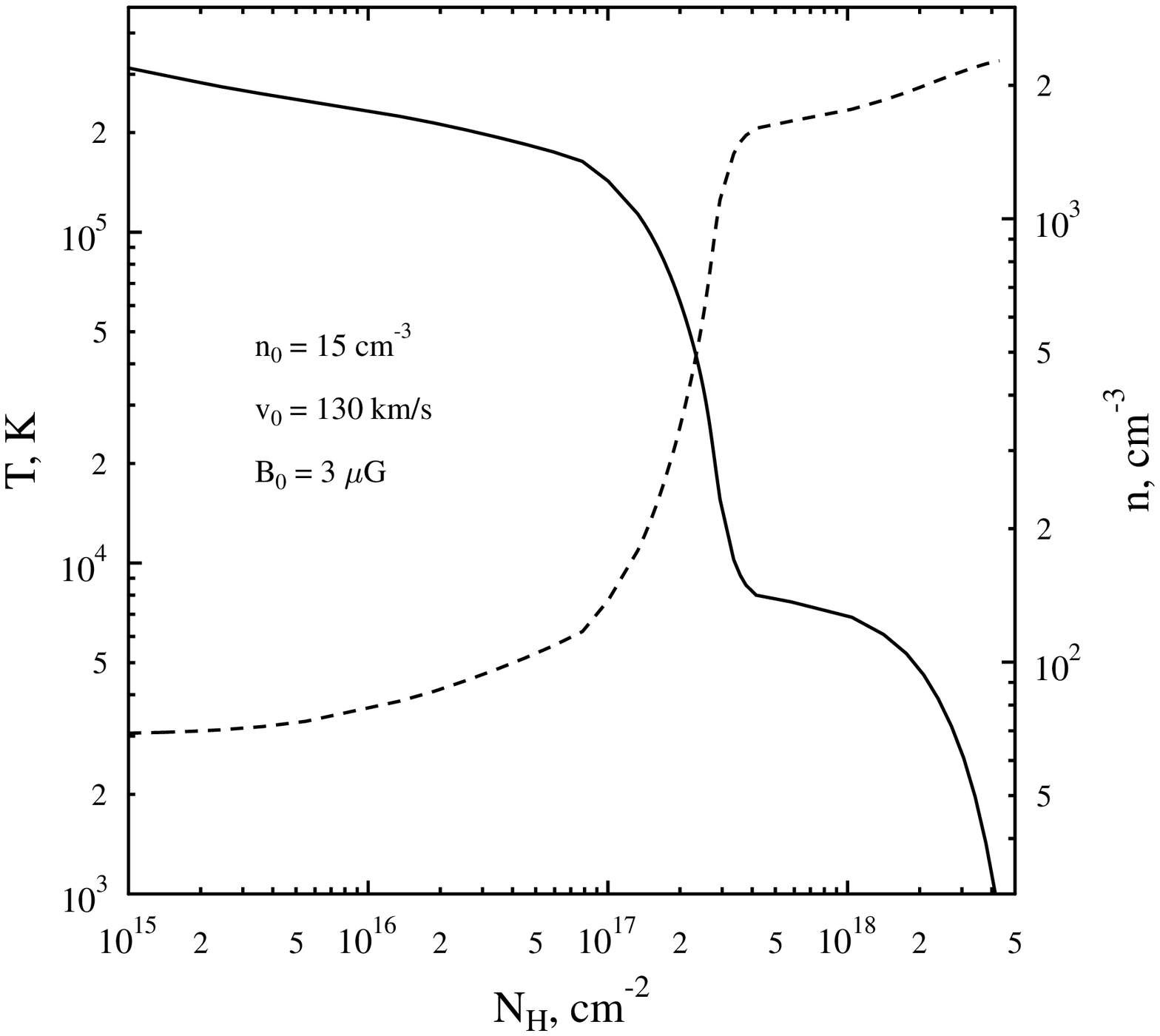}
\caption{Post-shock flow profile for shock and ISM parameters
plausible for the region of interaction of the north-eastern shell
of IC 443 SNR with an interstellar cloud. Here the upstream
density $n_0$ = 15 cm$^{-3}$, the shock velocity $v_0$ = 130 km/s,
and the perpendicular magnetic field $B_0$ = 3 $\mu$G.}
\label{fig:ic443}       
\end{figure*}

\begin{figure*}
\includegraphics[width=0.95\textwidth]{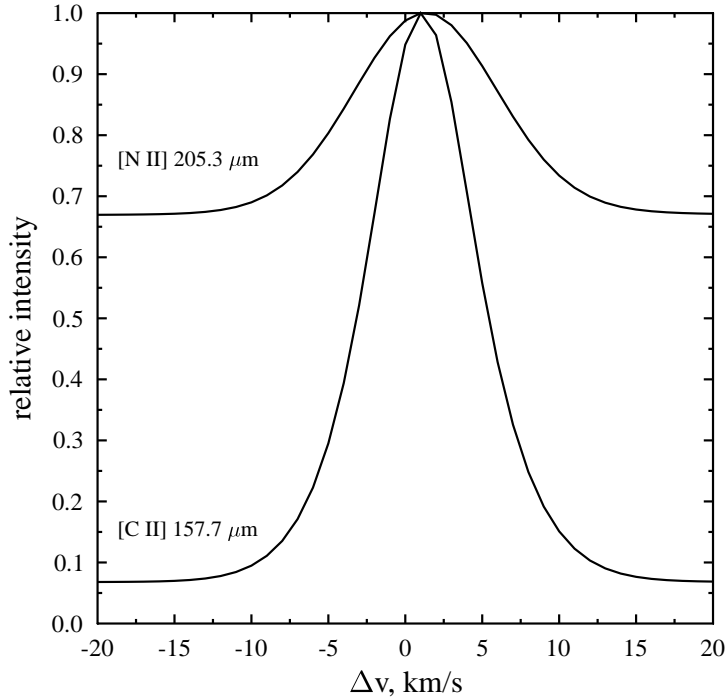}
\caption{ [C II] 157.7 $\mu$m and [N II] 205.3 $\mu$m line profiles
for the backgrounds of 140 MJy/sr and 130 MJy/sr, respectively,
estimated for the north-eastern shell of IC 443 SNR with Herschel
Spot (for the date 22/09/2010).}
\label{fig:lines}       
\end{figure*}

Figures~\ref{fig:ion} through \ref{fig:lines} and
Tables~\ref{table-lines-rel} and \ref{table-lines-abs} demonstrate
the results of modeling with SHELLS. The impact of efficient CR
acceleration on the shock is accounted for in all calculations by
using a reduced downstream temperature and the CR pressure in the
shock downstream.

The steady-state pre-shock ionization structure is shown in
Figure~\ref{fig:ion} for shocks with speeds of 100, 130, 200, and
310 $\kms$ illustrating also an effect of the CR fluid on the
pre-shock ionization. The plasma temperature and gas compression just
behind the CR modified shock affect the ionization structure. One
can see in Figure~\ref{fig:ion} that the ionization structure of 310
$\kms$ shock modified by CRs may be similar to that of 100 $\kms$
shock without CRs fluid effect. This Figure shows the ionization
level of hydrogen and helium; however, all abundant chemical species
up to Fe are tracked in the calculation. At speeds of approximately
130 km/s and above, hydrogen upstream becomes completely ionized,
and the extent of the radiative precursor is determined by the
competition between hydrogen and helium recombination and
photoionization.

Figure~\ref{fig:post-shock} illustrates the downstream cooling region of a 200 km/s shock.
Three lines correspond to three models with different assumptions regarding the efficiency of
CR acceleration. For efficient acceleration, energy conservation requires greater compression ratio
and lower downstream temperature. The three cases shown here illustrate
shocks with compression ratios of 4.6, 5.3 and 6.2.

Shown in Figure~\ref{fig:ic443} is the result of a calculation with shock and ISM parameters plausible for
the region of interaction of the north-eastern shell of the SNR~IC~443 with an interstellar cloud.
Unperturbed gas density is $n_0$ = 15~cm$^{-3}$, shock velocity $v_0$ = 130~km/s, and perpendicular
magnetic field $B_0$ = 3~$\mu$G are assumed.
The temperature and total density of the plasma are plotted as a function of gas column density from the subshock.
These quantities are then used to calculate the radiative transfer of emission lines in the cooling flow, which can be used for shock diagnostics.
Figure~\ref{fig:lines} shows the profiles of two of the most prominent infrared lines (with respect
to estimated backgrounds) for this case: C~II~157.7~$\mu$m and N~II~205.3~$\mu$m.

The model presented in this Section may be used to diagnose various
parameters of radiative shocks with observed infrared lines,
including the cosmic ray acceleration efficiency.

\section{Formation of CR spectral breaks in partly ionized shock precursors}
\label{sec:breaks}

\subsection{Gamma-ray observations}
\label{sec:gamma-obs}

The neutral component of the interstellar medium
may manifest itself in the gamma-ray emission of some collisionless shocks.

The recent Fermi-LAT observations of the so-called molecular SNRs
W44 and IC~443 \citep{Abdo10W44full, AbdoIC443_10, ackerman_ea_13}
indicate that the spectra of the gamma-ray producing protons
(integrated over the emission region) are typically steeper than the
DSA predictions for the spectra of the CRs confined in the
acceleration region. The steep photon spectra has been found in the
high energy gamma-ray spectra of some other remnants measured by,
e.g., the CANGAROO \citep{Enomoto02}, H.E.S.S \citep{Ahar06RXJ} and
MAGIC \citep{MAGICW51C11} atmospheric Cherenkov telescopes.

Particles accelerated by DSA mechanism generally comprised two
distinct populations --  CRs confined in the accelerator and the CRs
escaping the system and these two populations have very different
spectral shapes. The observed gamma-ray emission is a sum of the two
contributions \citep[see, e.g.,][]{eb11}. The common feature of the
molecular SNRs is a significant amount of dense molecular gas in
their surroundings. It has been argued on this ground \citep{MDS05},
that when a SNR interacts with a dense molecular cloud complex, the
conditions for particle confinement to the shock are different from
those adopted in conventional DSA modeling. Since the propagation of
resonant Alfv\'{e}n waves is inhibited by ion-neutral collisions,
some particles are not confined and so escape the emission volume
\citep[e.g.][]{DruryNeutral96, BykovMC00, MDS05}. The accelerated
particle partial escape should result in a spectral break in their
spectrum and thus, in that of the observed gamma emission. In a
clear-cut case of such a limited CR confinement, the spectral index
at the break should change by exactly one power $\Delta q=1$ due to
an effective reduction of the particle momentum space dimensionality
by one, since particles are confined in coordinate space only when
they are within a slab in momentum space oriented perpendicular to
the local magnetic field.

The most convincing evidence for the breaks of index one are the
recent Fermi-LAT and AGILE observations of W44
\citep{Abdo10W44full,Uchiyama10,AgileW44_11} (re-analyzed in
\citep{MDS_11NatCo}) and MAGIC observations of the SNR W51C
\citep{MAGICW51C11,MagicW51_12} These observations are encouraging
in that they unambiguously confirm the breaks. They demonstrate
departure from the traditional DSA models in fully ionized plasma
where the proton spectrum is a single power law with an exponential
cutoff. A possible explanation to the spectral break in the
gamma-ray emission of these objects is that neutral atoms modify the
nature of wave-particle interactions, leading to a spectral break.
This idea is discussed below.

\subsection{The DSA mechanism}

A great deal of the success of the diffusive shock acceleration
mechanism (DSA) is due to its ability to reproduce the observed
power law spectra associated with possible CR accelerators, such as
SNR shocks. For a classical step-like shock, propagating into an
ionized medium that supports sufficient CR scattering on both sides
of the shock, the mechanism predicts an almost universal power-law
energy spectrum for the accelerated CRs, $\propto E^{-q},$ with an
index $q=\left(r+2\right)/\left(r-1\right)$. It thus depends only on
the shock compression $r$, that is, however, close to four for
strong shocks. The backreaction of the accelerated CRs on the shock
structure may change this result noticeably, but not dramatically.
For strong shocks, $\mathcal{M}\gg1$, the spectrum hardens from
$q=2$ to $q\simeq1.5$, at most. This change is largely due to an
enhanced compression of the shock that results from the CR escape
flux and the reduction of the adiabatic index of the
CR/thermal-plasma mixture compared to the ordinary plasma (i.e.,
$\gamma=5/3\to4/3$). More problematic is to soften the CR spectrum
by their backreaction effects on the shock environment. This is
intuitively understandable, since spectrum steepening should
diminish the backreaction coming from the softening of the equation
of state ($\gamma\to4/3$) and, especially, from the CR escape flux.

The recent Fermi-LAT observations of the SNRs W44 and IC~443
\citep{Abdo10W44full,AbdoIC443_10} also urge a second look at the
DSA mechanism (see Section~\ref{sec:gamma-obs}).

\subsection{Mechanism for a spectral break}

When a SNR shock approaches a molecular cloud (MC) or a
pre-supernova swept-up shell with a significant amount of neutrals,
confinement of accelerated particles deteriorates. Due to the
particle interaction with magnetic fluctuations, confinement
requires their scales to be similar to the particle gyroradii
\citep{Drury83,blandfordeichler}. However, strong ion-neutral collisions
make the wave-particle interactions more sensitive to the particle
pitch angle, which can be understood from the following
consideration.

While the waves are in a strongly ionized (e.g., closer to the
shock) medium they propagate freely in a broad frequency range at
the Alfv\'{e}n speed $V_{A}=B/\sqrt{4\pi\rho_{i}}$ with the
frequencies $\omega=kV_{A}$. Here $k$ is the wave number (assumed
parallel to the local field $\mathbf{B}$) and $\rho_{i}$ is the ion
mass density. As long as the Alfv\'{e}n wave frequency is higher
than the ion-neutral collision frequency $\nu_{in}$, the waves are
only weakly damped. When, on the other hand, the ion-neutral
collision frequency is higher (deeper into the cloud), neutrals are
entrained by the oscillating plasma and the Alfv\'{e}n waves are
also able to propagate, albeit with a factor
$\sqrt{\rho_{i}/\rho_{0}}<1$ lower speed, where $\rho_{0}$ is the
neutral density. The propagation speed reduction occurs because
every ion is now ``loaded'' with $\rho_{0}/\rho_{i}$ neutrals. Now,
between these two regimes Alfv\'{e}n waves are heavily damped and
even disappear altogether for sufficiently small
$\rho_{i}/\rho_{0}\ll0.1$. The evanescence wave number range is then
bounded by $k_{1}=\nu_{in}/2V_{A}$ and
$k_{2}=2\sqrt{\rho_{i}/\rho_{0}}\nu_{in}/V_{A}$. These phenomena
have been studied in detail in \citep{KulsrNeutr69,ZweibelShull82},
and specifically in the context of the DSA in
\citep{VoelkNeutrDamp81,DruryNeutral96,BykovMC00,RevilleNeutr08}.
Now we turn to their impact on the particle confinement and
emissivity.

In the framework of a quasilinear wave-particle interaction, the
wave number $k$ is approximately related to the parallel (to the
magnetic field) component of the particle momentum $p_{\parallel}$
by the cyclotron resonance condition
$kp_{\parallel}/m=\pm\omega_{c}$, where the (non-relativistic)
gyro-frequency $\omega_{c}=eB/mc$. Note that the appearance of
$p_{\parallel}=p\mu$, where $\mu$ is the cosine of the pitch angle
instead of the often used ``sharpened'' \citep{Skill75a} resonance
condition $kp/m=\pm\omega_{c}$ is absolutely critical for the break
mechanism. The frequency range where the waves cannot propagate may
be conveniently translated into the parallel momentum range

\begin{equation}
p_{1}<\left|p_{\parallel}\right|<p_{2},\label{eq:ineq}
\end{equation}
with

\begin{equation}
p_{1}=2V_{A}m\omega_{c}/\nu_{in},\;\;
p_{2}=\frac{p_{1}}{4}\sqrt{\rho_{0}/\rho_{i}}>p_{1}.\label{eq:p12}
\end{equation}
That a spectral break must form at the photon energy corresponding
to the particle momentum $p=p_{1}=p_{br}$, can be readily explained
as follows. The 'dead zones'
$p_{1}<\left|p_{\parallel}\right|<p_{2}$ imply that particles with
$\left|p_{\parallel}\right|>p_{1}$ do not turn around (while moving
along the magnetic field) and escape from the region of CR-dense gas
collisions at a $p_{\parallel}/p$ fraction of the speed of light.
More specifically, particles with
$p_{1}<\left|p_{\parallel}\right|<p_{2}$ escape because they are not
scattered, whereas particles with
$\left|p_{\parallel}\right|>p_{2}$, because they maintain the sign
of $p_{\parallel}$, even though they scatter but cannot jump over
the gap $p_{1}<\left|p_{\parallel}\right|<p_{2}$. An exception to
this rule are particles with sufficiently large $p_{\perp}$ that can
be mirrored across the gap or overcome it via the resonance
broadening.

\begin{figure}
\includegraphics[clip,angle=-90,scale=0.5]{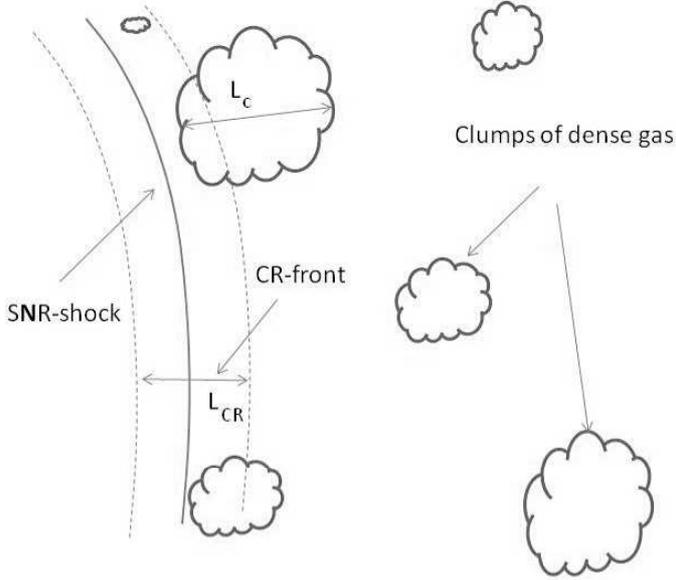}
\caption{\textbf{SNR shock propagating into dense gas environment}.
The filling factor of the gas clumps is small, while some of them
may be larger than the thickness of the CR layer near the shock
front.\label{fig:SNR-shock-propagating} }
\end{figure}

The break can also be explained in terms of the confinement times of
different groups of particles introduced above if we assume a low
density pre-shock medium with clumps of dense, partially neutral, gas.
Particles with
$\left|p_{\parallel}\right|>p_{1}$ spend only short time $\tau_{{\rm
esc}}\sim L_{{\rm c}}/c$ (where $L_{{\rm c}}$ is the size of the
clump) inside the gas clumps. They propagate ballistically and their
scattering time is assumed to be infinite, as there are no waves
they can interact with resonantly
($p_{1}<\left|p_{\parallel}\right|<p_{2}$) or they cannot change
their propagation direction ($\left|p_{\parallel}\right|>p_{2}$).
Particles with $\left|p_{\parallel}\right|<p_{1}$ are, on the
contrary, scattered intensively in pitch angle, they frequently
change their direction, and so sit in the clump for $\tau_{{\rm
conf}}\sim L_{{\rm c}}^{2}/\kappa\sim L_{{\rm
c}}^{2}/c^{2}\tau_{{\rm sc}}$. Here $\tau_{{\rm sc}}$ is their
pitch-angle scattering time and $\kappa$ is the associated diffusion
coefficient. Not only $\tau_{{\rm conf}}\gg\tau_{{\rm esc}}$ is
required, i.e., $\tau_{{\rm sc}}\ll L_{{\rm c}}/c$, but also
$\tau_{{\rm conf}}>L_{{\rm c}}/U_{{\rm sh}}$, which means that the
shock precursor is shorter than the clump $\kappa/U_{{\rm sh}}\lsim
L_{{\rm CR}}<L_{{\rm c}}$ (here $U_{{\rm sh}}$ is the shock
velocity, and $L_{{\rm CR}}$ is the thickness of the CR front near
the shock). The last condition ensures that particles with
$p_{\parallel}>p_{1}$ that escape through the clump after having
entered it from the shock side, will not interact with the shock
after they exit through the opposite side of the clump, thus
escaping upstream, Fig.\ref{fig:SNR-shock-propagating}. The reason
for that is a low level of Alfv\'{e}n wave turbulence ahead of the CR
precursor. We also assume that the ambient magnetic field does not
deviate strongly from the shock normal, in order to allow these
particles to escape through the far side of the clump.

While particles with $p>p_{1}$ escape from the regions of enhanced
gamma radiation (high gas density), an initially isotropic
distribution of accelerated particles is maintained only in a slab
in momentum space $\left|p_{\parallel}\right|<p_{1}$ and becomes
thus highly anisotropic (a 'pancake' distribution). What matters for
the integral emission, however, is a locally isotropic component
$\overline{f}$ of this new proton distribution. It can be introduced
by re-averaging the 'pancake' ($\left|p_{\parallel}\right|<p_{1}$)
distribution in pitch angle,
$\overline{f}\left(p\right)\equiv\intop_{0}^{1}f\left(p,\mu\right)d\mu$,
and is readily obtained assuming that particles remaining in the
dense gas (those with $\left|p_{\parallel}\right|<p_{1}$) maintain
the flat pitch-angle distribution, i.e.

\begin{eqnarray}
\overline{f}\left(p\right) & = &
\intop_{0}^{\mu_{1}}f_{0}\left(p\right)d\mu=\left\{
\begin{array}{cc}
\left(p_{1}/p\right)f_{0}\left(p\right), & p\ge p_{1}\\
f_{0}\left(p\right), & p<p_{1}
\end{array}\right.\label{eq:fbar}
\end{eqnarray}
where $f_{0}\left(p\right)$ is the initial (isotropic) distribution
function and $\mu_{1}=\min\left\{ p_{1}/p,1\right\} $. Thus, the
slope of the particle momentum distribution becomes steeper by
exactly one power above $p=p_{1}\equiv p_{{\rm br}}$. In particular,
any power-law distribution $\propto p^{-q}$, upon entering an MC,
turns into $p^{-q-1}$ at $p\ge p_{{\rm br}}$, and preserves its form
at $p<p_{{\rm br}}$.

Note that the broken power-law spectrum can only be maintained if
the filling factor $f_{{\rm gas}}$ of the dense gas with the
significant wave evanescence interval $\left(p_{1},p_{2}\right)$ is
relatively small, $f_{{\rm gas}}\ll1$, so that the overall particle
confinement and thus the acceleration are not strongly affected. If,
on the contrary, $f_{{\rm gas}}\sim1$, the resonant particles would
leak into the $(p_{1},p_{2})$ gap and escape from the accelerator in
large amounts, thus suppressing the acceleration. Note that
particles with sufficiently high momenta $p>p_{2}B_{0}/\delta B$,
where $\delta B/B_{0}$ is the effective mirror ratio of magnetic
perturbations, can ``jump'' over the gap. The primary $p^{-q}$ slope
should then be restored for such particles. Recent MAGIC
observations of the SNR W51C \citep{MAGICW51C11,MagicW51_12} indeed
point at such spectrum recovery at higher energies. It should also
be noted, that the $\Delta q=1$ break index is a limiting case of
identical gas clumps. The integrated emission from an ensemble of
clumps with different $p_{1}$ and $p_{2}$ should result in a more
complex spectrum.

\subsection{Break momentum}

While the one power spectral break in the pitch-angle averaged
particle distribution seems to be a robust environmental signature
of a weakly ionized medium into which the accelerated particles
propagate, the break momentum remains uncertain. According to
eq.(\ref{eq:p12}), $p_{br}$ ($\equiv p_{1}$) depends on the magnetic
field strength and ion density as well as on the ion-neutral
collision rate $\nu_{in}=n_{0}\left\langle \sigma V\right\rangle $.
Here $\left\langle \sigma V\right\rangle $ is the product of the
collision cross-section and collision velocity averaged over the
thermal distribution. Using an approximation of
\citep{Draine_McKee_93,DruryNeutral96} for $\left\langle \sigma
V\right\rangle $, $p_{br}$ can be estimated as
\begin{equation}
p_{br}/mc\simeq10B_{\mu}^{2}T_{4}^{-0.4}n_{0}^{-1}n_{i}^{-1/2}.\label{eq:p1}
\end{equation}
Here the gas temperature $T_{4}$ is measured in the units of
$10^{4}K$, magnetic field $B_{\mu}$ -in microgauss, $n_{0}$ and
$n_{i}$ (number densities corresponding to the neutral/ion mass
densities $\rho_{0}$ and $\rho_{i}$) \--in $cm^{-3}$. Note that the
numerical coefficient in the last expression may vary depending on
the average ion and neutral masses and can be higher by a factor of
a few \citep{KulsrNeutr69,NakanoMC84} than the estimate in
eq.(\ref{eq:p1}) suggests. The remaining quantities in the last
formula are also known too poorly to make an accurate independent
prediction of the position of the break in the gamma ray emission
region. Those are the regions near the blast wave where complicated
physical processes unfold, as discussed earlier
\citep{Shull_McKee_79,Draine_McKee_93,BykovMC00}. Also important may
be the ionization by the low energy CRs accelerated at the blast
wave. However, as their diffusion length is shorter than that of the
particles with $p\gsim p_{br}$, we may assume that they do not reach
the MC. Pre-ionization by the UV photons can also be ignored for the
column density $N>10^{19}cm^{-2}$ ahead of the shock beyond which
they are absorbed \citep{Uchiyama10}.
\cite{Uchiyama10}, using the earlier data from \citep{Reach05} have
also analyzed the parameters involved in eq.(\ref{eq:p1}) and found
the above estimate of $p_{br}$ to be in a good agreement with the
spectral break position measured by the Fermi-LAT.
Nevertheless, we may run the argument in reverse and use the
Fermi observations \citep{Abdo10W44full} of the gamma-ray
spectrum of SNR W44 to determine the break momentum in the parent
particle spectrum and constrain the parameters in eq.(\ref{eq:p1}).
Since we also know the amount of the slope variation $\Delta q$, we
can calculate the full spectrum up to the cut-off energy.

It should also be noted that in reality the break at $p=p_{{\rm
br}}$ is not infinitely sharp for the following reasons. The break
momentum may change in space due to variations of the gas parameters
(eq.{[}\ref{eq:p1}{]}), the resonance broadening
\citep{Dupree66,Acht81a} near $p=p_{1}=p_{{\rm br}}$ (so that
particles with $p\sim p_{1}$ are still scattered, albeit weakly) and
other factors, such as the contribution of small gas clumps with
$L_{c}\ll L_{{\rm CR}}$, Fig.\ref{fig:SNR-shock-propagating}. The
small clumps are submerged in the CR front and the CRs that escape
from them are readily replenished. Note that this effect may
decrease the break index $\Delta q$. However, the conversion of the
parent proton spectrum into the observable gamma emission introduces
a significant smoothing of the break, so that even a sharply broken
proton spectrum produces a smooth gamma spectrum. It provides an
excellent fit to the Fermi gamma-ray data without an ad hoc proton break
smoothing adopted by the Fermi-team \citep{Abdo10W44full} to fit the
data.

\section{Particle and gamma-photon spectra in molecular cloud SNRs}
\label{sect:gammaSNR}

To calculate the particle spectra, we need to determine the degree
of nonlinear modification of the shock structure. In principle, it
can be calculated consistently, given the shock parameters and the
particle maximum momentum, $p_{max}$. In the case of a broken
spectrum, $p_{br}$ likely plays the role of $p_{max}$, as a momentum
where the dominant contribution to the pressure of accelerated
particles comes from, thus setting the scale of the modified shock
precursor. Note that in the conventional nonlinear (NL) acceleration
theory, the cut-off momentum $p_{max}$ plays this role, because the
nonlinear spectra are sufficiently flat so as to make the pressure
diverge with momentum, unlike the broken spectra.

One of the best documented gamma emission spectra comes from the SNR
W44, so we fit these data using the above mechanism of the spectral
break. The break in the photon spectrum is observed at about $2$
GeV, which places the break in the proton distribution at about
$p_{br}\simeq7GeV/c$ \citep{Abdo10W44full}. For the strength of the
break $\Delta q=1$, the spectrum above it is clearly pressure
converging, so that the shock structure and the spectrum may be
calculated using this break momentum as the point of the maximum in
the CR partial pressure. More specifically, once the break momentum
is set, one can use an analytic approach \citep{MDru01} for a
stationary nonlinear acceleration problem using $p_{br}$ as an input
parameter.

Apart from $p_{br}$, the nonlinear solution depends on a number of
other parameters, such as the injection rate of thermal particles
into acceleration, Mach number, the precursor heating rate and the
shock velocity $V_{s}$. Of these parameters the latter is known
reasonably well, $V_{s}\approx$ 300 km/s, the injection rate can be
either calculated analytically for the parallel shock geometry
\citep{MDru01}, or inferred from the simulations
\citep{SpitkovskyHybr12}, while the other parameters are still
difficult to ascertain. Fortunately, in sufficiently strong shocks
the solution either stays close to the test particle (TP) solution
(leaving the shock structure only weakly modified) or else it
transitions to a strongly modified NL-solution regime. The TP regime
typically establishes in cases of moderate Mach numbers, low
injection rates and low $p_{max}$ (now probably closer to $p_{br}$),
while the NL regime is unavoidable in the opposite part of the
parameter space.

In the TP regime the spectrum is close to a power-law with the
spectral index 2 throughout the supra-thermal energy range. In the
NL regime, however, the spectrum develops a concave form, starting
from a softer spectrum at the injection energy, with the index
$q\simeq(r_{s}+2)/(r_{s}-1)>2$, where $r_{s}<4$ is the sub-shock
compression ratio. Then it hardens, primarily in the region $p\sim
mc$, where both the partial pressure and diffusivity of protons
change their momentum dependence. The slope reaches its minimum at
the cut-off (break) energy, which, depending on the degree of
nonlinearity, can be as low as 1.5 or even somewhat lower if the
cut-off is abrupt. The question now is into which of these two
categories the W44 spectrum falls? Generally, in cases of low
maximum (or, equivalently, low spectral break $p_{br}\lsim10$)
momentum, the shock modification is weak, so the spectrum is more
likely to be in a slightly nonlinear, almost TP regime. On the other
hand, there is a putative indication from the electron radio
emission that their spectrum may be close to $q_{e}\approx1.75$,
which could be the signature of a moderately nonlinear acceleration
process. It should be remembered, however, that this is a global
index across the W44 remnant. There are resolved bright filaments
where a canonical $\alpha=-0.5$ spectrum, corresponding precisely to
the TP parent electron spectrum with $q_{e}=2,$ is observed
\citep{W44Radio07}. Moreover, there are regions with the positive
indices $\alpha\lsim0.4$ which cannot be indicative of a DSA process
without corrections for subsequent spectral transformations such as
an absorption by thermal electrons.
These regions
may
contribute to the overall spectral hardening discussed
above, thus mimicking the acceleration nonlinearity. Finally, secondary
electrons give rise to the flattening of the radio spectrum as well
\citep{Uchiyama10}.

\begin{figure}
\includegraphics[bb=0bp 0bp 792bp 612bp,clip,scale=0.5,angle=-90]{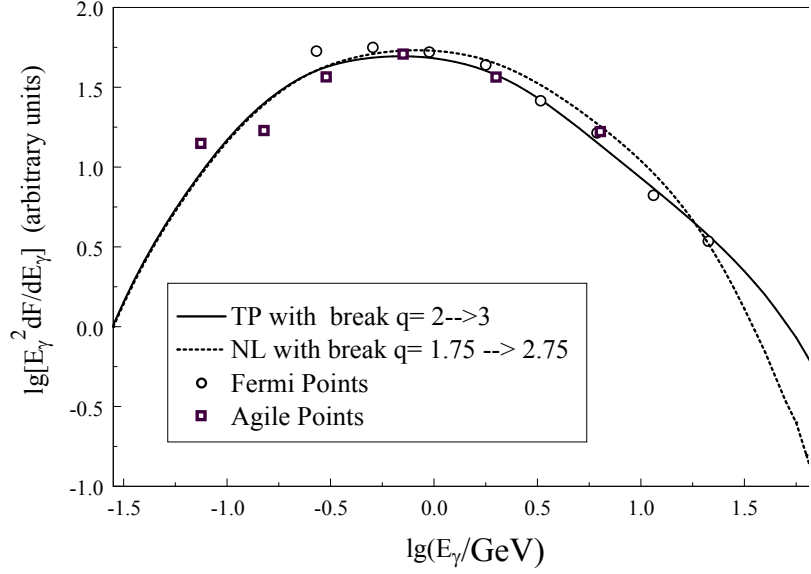}
\caption{\textbf{Gamma radiation spectra.} Photon spectra resulting
from $\pi^{0}$ decay and calculated for two different parent proton
spectra compared against the Fermi (circles) and AGILE (squares)
data. Solid line: a test particle acceleration regime with the
spectral index $q=2$ below the break and $q=3$ above the break at
$p_{br}=7GeV$/c. Dashed line: a moderately nonlinear acceleration
regime corresponding to the spectrum with $q\simeq1.75$ and
$q\simeq2.75$ below and above the break, respectively. Cut-offs are
placed at 300 GeV for TP- and 100 GeV, for NL-spectrum. Fermi and
AGILE data are adopted from \citep{Abdo10W44full,AgileW44_11},
respectively. Both curves are well within the error bars of Fermi
and AGILE (not shown for clarity), which, in turn, overlap
\citep{AgileW44_11}. \label{fig:Gamma-emission-from}}
\end{figure}

The above considerations somewhat weaken the radio data as a probe
for the slope of the electron and (more importantly) for the proton
spectrum. Therefore, the exact degree of nonlinearity of the
acceleration remains unknown and one can consider both the TP and
weakly NL regimes in calculations of the photon spectra, generated
in $p-p$ collisions. Specifically, \citet[][]{MDS05,MDS_11NatCo}
calculate the $\pi^{0}$ production rate and the gamma-ray
emissivity. In so doing, they adopt numerical recipe described in
detail in \citep{Kamae06,KarlssonKamae08}. The physical processes
behind these calculations are (i) collisions of accelerated protons
with the protons of the ambient gas resulting in $pp\to\pi^{0}$
reaction (ii) decay of $\pi^{0}$- mesons to generate an observable
gamma emission spectrum.

An example of such calculations is shown in
Fig.\ref{fig:Gamma-emission-from}. The best fit to the Fermi and
AGILE data is provided by a TP energy distribution ($\propto
E^{-2}$) below $p_{br}\simeq7$GeV/c with the spectrum steepening by
exactly one power above it. The spectrum steepening is perfectly
consistent with the proton partial escape described above (with no
parameters involved). For comparison, a weakly NL spectrum (guided
by the inferred electron spectrum with $q_{e}\approx1.75$, is also
used for these calculations (dashed line in
Fig.\ref{fig:Gamma-emission-from}), but its fit would require a
somewhat stronger break ($\Delta q\gsim1$) or a low momentum
cut-off, i.e. at least one additional free parameter. It is seen
that the mechanism for a break in the spectrum of shock accelerated
protons suggested in \citep{MDS05} provides a good fit to the recent
\citep{Abdo10W44full} Fermi-LAT and AGILE \citep{AgileW44_11}
observations of the SNR W44.  Of course, in assessing consistency of
the suggested spectral break phenomenon with the observed spectrum,
the errors in the data must be taken into account. The vertical
error bars near the break at 2 GeV, are fairly small (comparable to
the size of the symbols used to represent the data in
Fig.~\ref{fig:Gamma-emission-from}). More significant appears to be
the energy dispersion. However, in the most recent Fermi-LAT
publication (see \citet{ackerman_ea_13} including the Supplementary
Online Materials) the energy dispersion is estimated to be less than
5\% for these energies, so that the broken power law is indeed
consistent with the data.

Generally, spectral breaks offer a possible resolution to the well
known problem that some nonlinear DSA models produce spectra which
are considerably harder than a simple test particle spectrum, and
these are not consistent with the gamma-ray observations of some of
supernova remnants. However, the nonlinear spectrum -- i.e.,
diverging in energy-- exhausts the shock energy available for the
acceleration as the cut-off momentum grows, so that a broken
spectrum should form. Broken spectra are commonly observed and the
old paradigm of a single power-law with an exponential upper cut-off
is maladapted to the recent, greatly improved observations
\citep{AbdoIC443_10,Abdo10W44full}. Note, that the spectrum of the
RX J1713.7-3946 \citep{Ahar06RXJ} is also formally consistent with
the environmental break mechanism presumably operating in W44
surrounding but with a higher $p_{br}\sim10^{3}GeV/c$ and thus with
stronger acceleration nonlinearity \citep{MDS05}. However, this
remnant expands into a rather complicated environment, so it is
difficult to make the case for hadronic origin of the gamma-ray
emission \citep{AharNat04,Ahar06RXJ,WaxmanRXJ08,ellisonea12}. The
important role of the W44 remnant for the problem of CR origin is
that this particular remnant seems to be unlikely dominated by the
lepton emission due to Bremsstrahlung and inverse Compton scattering
\citep{Abdo10W44full,Uchiyama10} thus favoring the hadronic origin
of the gamma emission and bolstering the case for the SNR origin of
galactic CRs.

\section{Summary}
\label{sect:summ}

Collisionless shocks are ubiquitous in astrophysical objects and
are observed at all scales starting from the heliosphere and up
to cosmological scale shocks observed in clusters of galaxies.
Theoretical modeling of these shocks is challenging
because shock relaxation process involves collective plasma
oscillations producing long-lived highly nonequilibrium components.
Straightforward numerical simulations are an extremely resource
demanding task because of the very wide dynamical range of scales
and time required to the resolve all of the long-lived components.
Such task is even more difficult if neutral atoms and molecules
are significant in the partially ionized media.

In this brief review we have discussed some observational appearance
of collisionless shocks in partially ionized plasma,
described the most important physical processes
operating in the shocks and outlined the observational
perspective of nonthermal components diagnostics of astrophysical
collisionless shocks via multiwavelength observations.

Nonthermal components, i.e., energetic charged and neutral particles
and fluctuating magnetic fields, can drastically modify
the structure of the shock upstream providing both deceleration of
the plasma flow and also efficient heating of ions and electrons.
We argue that the processes of turbulence amplification and damping
observed in the heliosphere may help to understand the microphysics
of ion and electron heating by cosmic ray driven turbulence
in the upstream regions of large scale collisionless shocks observed
in the Galaxy and in clusters of galaxies. In turn, radiative signatures
of the astrophysical shocks can be a unique way to study microscopic
phenomena that can not be studied in the laboratory plasma on the Earth.

H$\alpha$ line diagnostics of Balmer shocks provides valuable information on
charge-exchange processes of the neutrals with both thermal and non-maxwellian
plasma components in the shock downstream.
Nonthermal components may reduce the plasma temperature comparing to what is
expected from the standard single-fluid Rankine-Hugoniot jump conditions
for a particular shock velocity. In the case of radiative shocks this reduction
would substantially modify the emission line spectrum coming from shock
downstream and thus may serve as a valuable diagnostic tool for fast shocks
interacting with clouds.
The effect of neutrals on the MHD wave damping in the upstream of supernova
shocks interacting with a molecular cloud may explain some spectral features
in GeV energy regime, recently revealed by {\sl Fermi} observations.

\begin{acknowledgements}
A.M.B., J.C.R. and M.A.M. thank Andre Balogh and the ISSI staff for
providing an inspiring atmosphere at the International Space Science
Institute Workshop in Bern in 2012, which has led to the new
collaboration. The authors thank the referee for a constructive
report. A.M.B. and A.M.K. acknowledge support from the RAS Programs
P~21 and OFN~16, and from the Ministry of Education and Science of
Russian Federation (Agreement No. 8409, 2012). They performed the
simulations at the Joint Supercomputing Centre (JSCC RAS) and the
Supercomputing Centre at Ioffe Institute, St. Petersburg. M.M.
acknowledges support by the Department of Energy, Grant No.
DE-FG02-04ER54738. A.E.V. acknowledges support by NASA grants
NNX09AC15G and NNX12AO73G.
\end{acknowledgements}


\bibliographystyle{svjour}


\bibliography{rsh}
\end{document}